\documentclass[aps,prb, twocolumn, nofootinbib, superscriptaddress]{revtex4-1}
\usepackage{amsmath}
\usepackage{amssymb}
\usepackage{graphicx}
\usepackage{latexsym}
\usepackage{amsfonts}
\usepackage{layout}
\usepackage{verbatim}
\usepackage{dcolumn}    % Align table columns on decimal point
\usepackage{bm}         % bold math
\usepackage[T1]{fontenc}%..........Type 1 outline fonts
\usepackage{bbold} 
\usepackage{mathrsfs}
\usepackage{mathtools}
\usepackage{graphicx}
\usepackage{tikz}
\usepackage{lipsum}
\usepackage{pgf}
\usepackage{float}
\usepackage{xcolor}
\usepackage{hyperref}
\usepackage{tabularx}
\usepackage{cancel}
\hypersetup{colorlinks=true,linkcolor={red!50!black},citecolor={blue!50!black},urlcolor={blue!80!black}}
\usepackage[caption=false]{subfig}
\usepackage{float}
\begin{document}
%========================================================================
%========================================================================
\title{Electrical operation of planar Ge hole spin qubits in an in-plane magnetic field}

\author{Abhikbrata Sarkar}
\affiliation{School of Physics, The University of New South Wales, Sydney 2052, Australia}

\author{Zhanning Wang}
\affiliation{School of Physics, The University of New South Wales, Sydney 2052, Australia}

\author{Matthew Rendell}
\affiliation{School of Physics, The University of New South Wales, Sydney 2052, Australia}

\author{Nico W. Hendrickx}
\affiliation{QuTech and Kavli Institute of
Nanoscience, Delft University of Technology, 2628 CJ Delft,
The Netherlands}

\author{Menno Veldhorst}
\affiliation{QuTech and Kavli Institute of
Nanoscience, Delft University of Technology, 2628 CJ Delft,
The Netherlands}

\author{Giordano Scappucci}
\affiliation{QuTech and Kavli Institute of
Nanoscience, Delft University of Technology, 2628 CJ Delft,
The Netherlands}

\author{Mohammad Khalifa}
\affiliation{Department of Electrical and Computer Engineering. University of British Columbia, Vancouver, B.C., Canada.}
\affiliation{Quantum Matter Institute, University of British Columbia, Vancouver, B.C., Canada.}

\author{Joe Salfi}
\affiliation{Department of Electrical and Computer Engineering. University of British Columbia, Vancouver, B.C., Canada.}
\affiliation{Quantum Matter Institute, University of British Columbia, Vancouver, B.C., Canada.}

\author{Andre Saraiva}
\affiliation{School of Electrical Engineering and Telecommunications, The University of New South Wales, Sydney 2052, Australia}

\author{A. S. Dzurak}
\affiliation{School of Electrical Engineering and Telecommunications, The University of New South Wales, Sydney 2052, Australia}

\author{A. R. Hamilton}
\affiliation{School of Physics, The University of New South Wales, Sydney 2052, Australia}

\author{Dimitrie Culcer}
\affiliation{School of Physics, The University of New South Wales, Sydney 2052, Australia}

\date{\today}
%========================================================================
%========================================================================
\begin{abstract}
Hole spin qubits in group-IV semiconductors, especially Ge and Si, are actively investigated as platforms for ultrafast electrical spin manipulation thanks to their strong spin-orbit coupling. Nevertheless, the theoretical understanding of spin dynamics in these systems is in the early stage of development, particularly for in-plane magnetic fields as used in the vast majority of experiments. In this work we present a comprehensive theory of spin physics in planar Ge hole quantum dots in an in-plane magnetic field, where the orbital terms play a dominant role in qubit physics, and provide a brief comparison with experimental measurements of the angular dependence of electrically driven spin resonance. We focus the theoretical analysis on electrical spin operation, phonon-induced relaxation, and the existence of coherence sweet spots. We find that the choice of magnetic field orientation makes a substantial difference for the properties of hole spin qubits. Furthermore, although the Schrieffer-Wolff approximation can describe electron dipole spin resonance (EDSR), it does not capture the fundamental spin dynamics underlying qubit coherence. Specifically, we find that: (i) EDSR for in-plane magnetic fields varies non-linearly with the field strength and weaker than for perpendicular magnetic fields; (ii) The EDSR Rabi frequency is maximized when the a.c. electric field is aligned parallel to the magnetic field, and vanishes when the two are perpendicular; % (iii) The Rabi ratio $T_1/T_\pi$ i.e. the number of EDSR gate operation per unit relaxation time, is expected to be as large as $5{\times}10^5$ at the magnetic fields used experimentally; 
(iii) The orbital magnetic field terms make the in-plane $g$-factor strongly anisotropic in a squeezed dot, in excellent agreement with experimental measurements; (iv) Focusing on random telegraph noise, we show that coherence sweet spots do not exist in an in-plane magnetic field, as the orbital magnetic field terms expose the qubit to all components of the defect electric field. These findings will provide a guideline for experiments to design ultrafast, highly coherent hole spin qubits in Ge. 
\end{abstract}
\maketitle

\section{Introduction}

% Schrieffer-Wolff works for EDSR but not for orbital terms. 
% Turn off orbital terms T2* increases by two orders of magnitude.

% Nico main points
% Harmonise T2* with Nico's experiment

% Menno main points

Solid state spin qubits are prime candidates for scalable, highly coherent quantum computing platforms. \cite{kane1998silicon, loss1998quantum, petta2005coherent, Hanson2007, Hanson2008, Zwanenburg2013, chatterjee2021semiconductor, scappucci2021germanium, fang2022recent} Among these group IV materials such as Ge and Si stand out thanks to the absence of piezo-electric interaction with phonons \cite{cardona2005fundamentals} and the possibility of isotopic purification, which eliminates the contact hyperfine coupling to the nuclear field,\cite{itoh1993ge, itoh2003si} with the maturity of semiconductor microfabrication as an added advantage. Recent years have witnessed a concerted push towards all-electrical qubit operation, since electric fields are easier to apply and localise than magnetic fields, and electrically operated qubit gates offer significant improvements in speed and power consumption as compared to magnetic gates. A series of theoretical predictions\cite{bulaev2007electric, kloeffel2011strong, kloeffel2013circuit, chesi2014controlling} as well as experimental leaps in growth techniques and sample quality \cite{dobbie2012ultra, sammak2019shallow, lodari2019light} have led to a surge in interest in spin-3/2 hole systems in group IV materials. The strong and multifaceted spin-orbit coupling experienced by holes, \cite{cardona2005fundamentals, winkler2003spin, Winkler2008, durnev2014spin, marcellina2017spin} their anisotropic and tunable $g$-tensor, \cite{DanneauPRL97, miserev2017dimensional, hung2017spin, Qvist2022Aniso-g, Abadillo2023Aniso-g} and the absence of a valley degree of freedom makes them ideal for electrical spin manipulation, with Ge offering the additional advantages of a small effective mass \cite{terrazos2021theory} and ease of ohmic contact formation. Compared to electrons, the weaker hyperfine coupling for holes\cite{keane2011nanolett, chekhovich2011direct} due to the absence of the contact interaction significantly reduces the nuclear field contribution to spin decoherence, while the hole spin-3/2 is responsible for physics with no counterpart in electron systems, \cite{Culcer_Precession_2006, Winkler2008, Hong2018, Abadillo2018Magic, Cullen_2021} which may offer flexibility in future design strategies --  for example magic angles have been predicted for acceptor qubits, \cite{Abadillo2018Magic} at which dipole-dipole entanglement can be switched off without switching off the electric dipole moments of single qubits. 

% Initial period: readout, self-assembled, many-hole
Remarkable progress on hole spin qubits in several architectures has spanned more than a decade, with an overwhelming focus on Ge and Si.\cite{chatterjee2021semiconductor, scappucci2021germanium, fang2022recent} Initial work focused on measuring hole spin states, \cite{roddaro2008PRL, zwanenburg2009spin, li2015pauli, liles2018spin} % Both of these are Si Nanowires - first few holes
relaxation and dephasing times, \cite{hu2012hole, higginboth2014nanolettt, vuku2018single}
% SiGe self-assembled
single spin electrical control, \cite{pribiag2013electrical} 
readout and control of the $g$-tensor \cite{ares2013sige, ares2013nature, brauns2016anisotropic, watzinger2016heavy, voisin2016electrical, srinivasan2016electrical, mizokuchi2018ballistic, marcellina2018electrical, wei2020estimation, zhang2021anisotropic, liles2021electrical}
and of spin states in multiple dots, \cite{li2015pauli,  bohuslavskyi2016pauli, wang2016PSB, van2018readout, Ezzouch2021SiDQD} and achieving single-spin spin qubits. \cite{maurand2016cmos, watzinger2018germanium} In recent years the development of strained germanium in SiGe heterostructures \cite{lodari2019light, lodari2021low, terrazos2021theory} provided a low-disorder environment, which supported the development of single-hole qubits, \cite{hendrickx2020single} singlet-triplet qubits, \cite{jirovec2021singlet} universal quantum logic, \cite{Hendrickx2020} and a four-qubit germanium quantum processor.\cite{hendrickx2021four} Experiments have demonstrated ultrafast spin manipulation using the spin-orbit interaction \cite{Hendrickx2020, hendrickx2020single, froning2021strong, wang2022ultrafast} and EDSR Rabi oscillations as fast as $540$ MHz, \cite{wang2022ultrafast}
electrical control of the underlying spin-orbit coupling \cite{gao2020site, Liu2022GateTunable} and charge sensing using a superconducting resonator, \cite{Ungerer2022SensingResonator}
while relaxation times of up to 32 ms have been measured in Ge dots. \cite{lawrie2020spin} Hole spins in Ge have been used as quantum simulators \cite{Wang2022GeSimulator} and control of an array of 16 Ge hole dots has been demonstrated. \cite{Borsoi2022-16QDs} The development of hybrid structures offers another path towards entanglement, with the demonstrations of superconductivity in planar Ge, \cite{hendrickx2018gate, Aggarwal2021EnhanceSC, valentini2023radio} hole coupling to a superconducting resonator, \cite{li2018coupling} dipole coupling to a microwave resonator, \cite{xu2020dipole} charge sensing using a superconducting resonator \cite{Ungerer2022SensingResonator}
and devices such as transistors and interferometers. \cite{vigneau2019germanium} Theoretically, the interplay of spin-orbit coupling and superconductivity in hybrid semiconductor-superconductor structures is only now being studied in the context of quantum computing. \cite{Lidal2023SNS}

Concomitantly, Si qubits have also registered remarkable recent progress, with coherence times of up to 10ms in Si:B acceptors \cite{kobayashi2021engineering} and the detection of sweet spots as a function of the top gate field in quantum dots, \cite{piot2022single} sweet spots having been the subject of a number of theoretical studies. \cite{kloeffel2011strong, salfi2016charge, Salfi2016AcceptorQC, Abadillo2018Magic, kloeffel2018direct, marcellina2018electrical, terrazos2021theory, wang2021optimal, bosco2021sweetspots, bosco2022hole, wang2022modelling, malkoc2022charge} Anisotropic exchange was also used to entangle two hole qubits,  \cite{Geyer2022Exchange} hole coupling to a superconducting resonator has been demonstrated, \cite{Maurand2022Photon}
and progress has been made towards higher-temperature operation with the observation of Coulomb diamonds up to 25K \cite{shimatani2020temperature} and single-qubit operation above 4K. \cite{Camenzind2022_4K} 

Despite experimental advances on many fronts, constructing an all-encompassing theory to describe hole physics in group-IV semiconductor quantum dots is challenging. In particular, owing to different effective masses and intrinsic spin-orbit gap in the valence band of these materials, the wide range of QD sizes as well as spin manipulation timescales exhibited in experiments render it difficult to provide a theory of spin qubits common to Si and Ge. In this work we will focus on Ge, which is somewhat more tractable analytically than Si, and describe electrical qubit operation in an in-plane magnetic field. This choice is motivated by the observation that experiments has overwhelmingly favoured in-plane magnetic fields, \cite{watzinger2018germanium, hendrickx2020single} partly because a transverse magnetic field makes it easier to suppress hyperfine and cyclotron quantization effects, and partly in order to avoid the strong orbital coupling of the perpendicular field, which can cause a large diamagnetic shift and also affect tunnel rates. For spin-3/2 holes, in-plane magnetic fields are highly non-trivial, because the spin and orbital degrees of freedom are intertwined. The in-plane $g$-factor is very small, and in Ge most of it comes from the octupole interaction with the magnetic field.\cite{winkler2004spin} Whereas in-plane magnetic fields have been considered for realistic devices in recent theoretical studies, \cite{ciocoiu2022towards, wang2022modelling, martinez2022hole} this has been largely from an engineering point of view, and the orbital effect of an in-plane magnetic field in planar quantum dots has largely been neglected, though they have been shown to play an important role in nanowires. \cite{adelsberger2022hole} It has thus not been possible to date to construct a full picture of spin dynamics and electrical spin operation in planar Ge hole dots in an in-plane magnetic field. This has left several outstanding questions unanswered: \textit{What determines the speed of EDSR, as well as the relaxation time $T_1$? Is there an optimal magnetic field orientation for driving a spin-orbit qubit? Do coherence sweet spots exist when the magnetic field is in the plane?}
%added "spin-orbit", because I think the physics of optimal orientation of the magnetic field comes from the one qubit calculation -- Abhik.
\begin{figure}[tbp]
\includegraphics[width=0.48\textwidth]{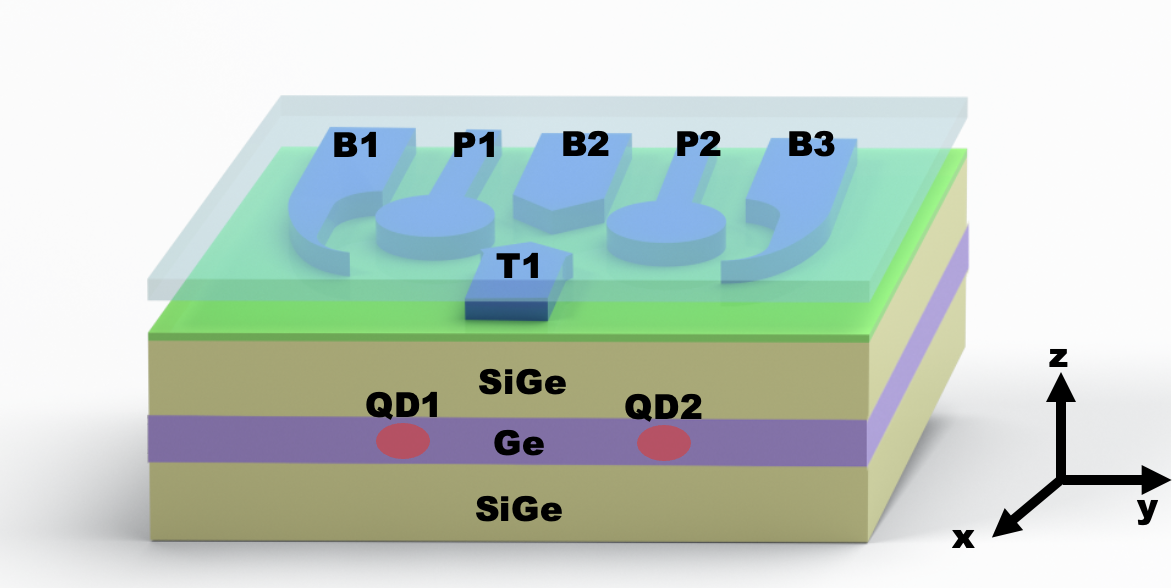}
\caption{\label{fig: Germanium Qubit} \textbf{A prototype double quantum dot in a strained germanium hole system.} The strained quantum well is grown epitaxially on a strain-relaxed SiGe layer. Gate B2 and T1 control the inter-dot tunneling. The growth direction is $\hat{\bm z}$.}
\end{figure}

In this work we seek to answer these questions. Our focus will be on gate-defined Ge quantum dots, with the parent 2DHG exhibiting very high mobility\cite{sammak2019shallow, lodari2022lightly}, low percolation density,\cite{lodari2021low,stehouwer2023germanium}, and a low effective mass of $m_h{=}\,0.05\,m_e;$\cite{hendrickx2018gate,lodari2019light}, which all aid the formation of quantum dots. One reason for our choice of Ge is its band alignment, which makes it the only group IV material suitable for growing quantum wells. Another reason is pragmatic -- it is easier to describe theoretically. This is because the spin-orbit splitting in Ge ($\Delta_{SO}{=}325\,\text{meV}$) is stronger than that in Si ($\Delta_{SO}{=}44\,\text{meV}$), resulting in a large separation of the spin-orbit/split-off (SO) band from the heavy- and light-hole subspaces. This ensures that the $4\!\times\!4$ Luttinger Hamiltonian formalism for spin-$3/2$ is adequate for the topmost Ge valence band, as opposed to Si where the Rashba mediated electrical control features both heavy hole-light hole (HH-LH) coupling contributions and heavy hole-split off (HH-SO) coupling contributions. Ge has a noticeable cubic-symmetry contribution to the Luttinger Hamiltonian. It is strong enough to enable electrical spin manipulation in planar dots, \cite{terrazos2021theory, wang2021optimal} making Ge ideal for electrical spin operation, but not as strong as in Si, such that it can be treated perturbatively.

Figure.~\ref{fig: Germanium Qubit} provides a schematic of the device architecture for studying a planar germanium hole quantum dot qubit. In this paper we concentrate on developing the key formalisms for describing the spin physics of hole spin qubits in an in-plane magnetic field. To avoid unneccessary complexity, and to keep our results generally applicable, we avoid making the analysis overly device specific; therefore we do not consider effects of gate electrode induced non-uniform strain which can lead to a significant modification of the spin-orbit interaction;\cite{liles2021electrical, corley2023nanoscale} or Fowler-Norheim tunnelling of hole states through the SiGe barrier leading to charge accumulation at the interface between the semiconductor and the gate dielectric; or light-hole penetration through the SiGe barrier. Addressing these would require detailed finite element numerical simulation techniques on top of the theory developed here.

For an applied in-plane magnetic field operation of planar Ge hole QD, in the presence of uniaxial strain but neglecting shear strain, we show that: 
(i) EDSR is linear in the magnetic field with nonlinear corrections emerging at larger fields, and is driven by Rashba spin-orbit coupling rather than by the orbital magnetic field terms; the picture that emerges is that the orbital B terms give rise to the finite Zeeman splitting between the qubit energy levels, while the Rashba spin-orbit coupling gives rise to transitions between them. The EDSR Rabi frequency is a maximum when the electric field is parallel to the magnetic field, and vanishes when the two are perpendicular; it can be sizable despite the smallness of the in-plane $g$-factor, and has a non-monotonic dependence on $B$; (ii) The relaxation rate is due to bulk acoustic phonons and is proportional to $B^3$ in the leading power; \cite{li2020hole} (iii) For a squeezed (elliptical) dot with an aspect ratio $L_y/L_x{=}\,2$ the spin-flip frequency is an order of magnitude faster than for a circular dot allowing Rabi time $\sim\!10$ ns with $\sim\!10^5$ operations within time $T_1$; (iv) For a squeezed dot, due to the effect of the magnetic field vector potential terms, the in-plane $g$-factor exhibits a strong anisotropy resulting in oscillatory behavior as the magnetic field is rotated in the plane of the dot, an observation supported by new experimental measurements shown in Sec.~V of this paper; (v) Although extrema in the qubit Zeeman splitting as a function of the top gate voltage exist in the same way as for a perpendicular magnetic field,\cite{wang2021optimal} coherence sweet spots \textit{do not exist} for in-plane magnetic fields for noise induced by charge defects in the plane of the quantum well. The physics underlying this process is associated with the orbital magnetic field terms and is not captured by the Schrieffer-Wolff approximation (although electrical qubit operation including EDSR can be described in a simplified Schrieffer-Wolff picture). 

% DIMI - IMPORTANT PHYSICS POINT: ARE WE SIMPLY SAYING THAT TO BE PROTECTED AGAINST RTS, THE ELECTRIC FIELD OF THE RTS SHOULD NOT BE PARALLEL TO THE APPLIED MAGNETIC FIELD? 
% No, it is not that simple. Abhik ran the T2* calculation with E//y and B//x and T2* still increases monotonically as a function of F_z.

The outline of this paper is as follows. Section \ref{sec:Hamil} lays down the foundation of the semi-analytical model of a Ge hole quantum dot qubit withing the framework of effective mass theory. Next we discuss the properties of circularly symmetric dots in Sec.~\ref{sec:Zeeman}, including the qubit Zeeman splitting, EDSR, relaxation and dephasing. In Sec.~\ref{sec:ellipt} we focus on elliptical dots and study their EDSR and coherence properties, while in Sec.~\ref{experiment} we discuss the consequences of $g$-factor anisotropy and compare our predictions with recent experimental results. We end with a summary and outlook in Sec.~\ref{sec:outlook}.

\section{Hamiltonian and model}\label{sec:Hamil}

% Hole charge

The topmost valence band in Ge has orbital angular momentum $l{=}1$. When the hole spin $s{=}\frac{1}{2}$ is taken into account the resultant states at the $\mathbf{\Gamma}$-point are eigenstates of the total angular momentum $\mathbf{J}{=}(\mathbf{L}+\mathbf{S})$. The four-fold degenerate $j{=}\frac{3}{2}$ states are separated by the spin-orbit gap $\Delta_0$ from the $j{=}\frac{1}{2}$ two-fold degenerate split-off states. For Ge the spin-orbit gap $\Delta_0{=}325\,\text{meV}$, so the split-off band is safely disregarded in describing hole motion in the topmost valence bands. The $\left|\frac{3}{2}\right\rangle$ and $\left|-\frac{3}{2}\right\rangle$ states constitute the heavy-hole (HH) manifold while the $\left|\frac{1}{2}\right\rangle$ and $\left|-\frac{1}{2}\right\rangle$ states represent the light-hole (LH) manifold. The Luttinger Hamiltonian describes the hole motion in the topmost valence bands and has the following form in the $j{=}\frac{3}{2}$ basis $\left\{\left|\frac{3}{2}\right\rangle,\left|-\frac{3}{2}\right\rangle,\left|\frac{1}{2}\right\rangle,\left|-\frac{1}{2}\right\rangle\right\}$:
\begin{equation}\label{eq:LKHamiltonian}
    H_{LK}(k)=\begin{pmatrix}
    P'+Q' & 0 & L' & M' \\
    0 & P'+Q' & M'^* & -L'^* \\
    L'^* & M' & P'-Q' & 0 \\
    M'^* & -L' & 0 & P'-Q' \\
    \end{pmatrix}
\end{equation}
where the matrix elements of the Luttinger Hamiltonian comprise the $k$-dependent part and strain-induced perturbations: $P'{=}P(\mathbf{k})+P_\varepsilon,\,Q'{=}Q(\mathbf{k})+Q_\varepsilon,\,L'{=}L(\mathbf{k})+L_\varepsilon,\,M'{=}M(\mathbf{k})+M_\varepsilon$. The kinetic energy terms are $P\,{=}\frac{\hbar^2\gamma_1}{2m_0}(k_x^2+k_y^2+k_z^2)$, $Q\,{=}\frac{\hbar^2\gamma_2}{2m_0}(k_x^2+k_y^2-2k_z^2)$, $L\,{=}\frac{-\sqrt{3}\hbar^2\gamma_3}{m_0}(\{k_x,k_z\}-i \{k_y,k_z\})$ and $M\,{=}\frac{\sqrt{3}\hbar^2}{2m_0}\{-\gamma_2(k_x^2-k_y^2)+2i\gamma_3\{k_x,k_y\}\}$. The strain terms are: $P_\varepsilon{=}-a(\varepsilon_{xx}+\varepsilon_{yy}+\varepsilon_{zz})$, $Q_\varepsilon{=}-b(\varepsilon_{xx}+\varepsilon_{yy}-2\varepsilon_{zz})$, $L_\varepsilon{=}d(\varepsilon_{xz}-i \varepsilon_{yz})$ and $M_\varepsilon{=}\frac{\sqrt{3}}{2}b (\varepsilon_{xx}-\varepsilon_{yy})-d \varepsilon_{xy}$. Here $m_0$ is the bare electron mass while $\gamma_1$,$\gamma_2$ and $\gamma_3$ are Luttinger parameters. The constant $a{=}2\,\text{eV}$ is the hydrostatic deformation potential, $b{=}{-}2.16\,\text{eV}$ is the uniaxial deformation potential, and $d{=}{-}6.06\,\text{eV}$ accounts for the shear deformation potential. In most Ge/GeSi samples there is considerable strain in the quantum well, which significantly increases the splitting between light and heavy holes compared to silicon - here we take the compressive strain to be $0.6\%$.\cite{sammak2019shallow}
%added Zhanning's paper as the reference for 0.6% strain. Is that okay? -- Abhik.
The strain tensor components in the plane are: $\varepsilon_{xx}{=}\varepsilon_{yy}{=}-0.006$. The in-plane compressive strain elongates the out-of-plane lattice constant via $\varepsilon_{zz}{=}-2\frac{C_{12}}{C_{11}}\varepsilon_{xx}=0.004$; with $C_{12}{=}44$ GPa, $C_{11}{=}126$ GPa.\cite{wortman1965young}
%added the reference -- Abhik.
We assume the off-diagonal shear elements of the strain tensor to be $\varepsilon_{ij|_{i\neq j}}{=}0$. The out-of-plane confinement is described by a one-dimensional infinite square well potential
\begin{equation}
  V(z) =
    \begin{cases}
      \infty & z\in \left\{-\frac{L_z}{2},\frac{L_z}{2}\right\}\\
      0 & \text{otherwise}\\
    \end{cases}.       
\end{equation}
The coupling to the top-gate electric field, denoted by $F_z$, gives an additional term $eF_zz$ in the Hamiltonian. The in-plane confinement is modelled by a parabolic potential $V_{x,y}{=}\frac{1}{2}(\lambda_x^2x^2+\lambda_y^2y^2)$, where $\lambda_x,\,\lambda_y$ are determined by the dot dimensions $L_x,\,L_y$ in the plane. The effective hole QD Hamiltonian is given by:
\begin{equation}
    H_{0D}=H_{LK}(\mathbf{k})+eF_zz+V_{\text{conf}}, 
\end{equation}
where $V_{\text{conf}}=V_{x,y}+V(z)$ is the total confinement potential. The Zeeman interaction is given by,
\begin{equation}\label{eq:Zeeman}
    H_Z=-2\kappa\mu_B\mathbf{B}\cdot \mathbf{J}-2q\mu_B\mathbf{B}\cdot\mathcal{J},
\end{equation}
where $\mathbf{J}=\{J_x,\,J_y,\,J_z\}$, $\mathcal{J}=\{J_x^3,J_y^3,J_z^3\}$; and $J_x,\,J_y,\,J_z$ are the $4{\times}4$ Pauli matrices. The $\mathcal{J}$-terms are vital in order to obtain the correct in-plane $g$-factor $\approx 0.25$.

In the presence of a magnetic field, the canonical momentum of holes in topmost valence band becomes $\mathbf{k}\rightarrow\left(\mathbf{k}+\frac{e\mathbf{A}}{\hbar}\right)$. The hole spin in a planar Ge quantum dot is then described by
\begin{equation}\label{eq:totalQDH}
H_{QD}=H_{LK}\!\!\left(\mathbf{k}+\frac{e\mathbf{A}}{\hbar}\right)+eF_zz+V_{\text{conf}}+H_Z. 
\end{equation}
The resultant modifications in $H_{LK}\left(\mathbf{k}+\frac{e\mathbf{A}}{\hbar}\right)$ lead to non-trivial contributions to the effective spin-orbit interaction in the HH and LH manifolds, as well as between them. To check the gauge invariance of our theoretical framework, we consistently diagonalise the effective quantum dot Hamiltonian using two different gauges:
\begin{enumerate}
    \item $\mathbf{A}{=}-\frac{1}{2}B_z y\mathbf{\hat{e}_x}+\frac{1}{2}B_z x\mathbf{\hat{e}_y}+(B_x y-B_y x)\mathbf{\hat{e}_z}$.
    \item The symmetric gauge: $\mathbf{A} \, {=} \, \frac{1}{2} \, \mathbf{B}\times\mathbf{r} \, {=} \, \frac{1}{2}(B_yz-B_zy) \, \mathbf{\hat{e}_x}+\frac{1}{2}(B_zx-B_xz)\, \mathbf{\hat{e}_y}+\frac{1}{2}(B_x y-B_y x) \, \mathbf{\hat{e}_z}$.
\end{enumerate}
All calculations have been performed in both gauges, yielding consistent results. 

%mometum operator: self-adjoint extension
The eigenstates of the hole QD can be expressed as linear combinations of states belonging to a basis in which the \textit{bare} QD Hamiltonian is diagonal, i.e. $\left|\Psi_{QD}\right\rangle=\sum_i c_i\psi_i(x,y,z)\left|\phi_j\right\rangle$, where the \textit{bare} Hamiltonian refers to $H_{QD}$ of Eq.~\ref{eq:totalQDH} with its off-diagonal elements set to zero, following the practice of ${\bm k}\cdot{\bm p}$ theory, as well as the external magnetic field set to zero. We choose the spatial wave functions as $\psi(x,y,z){=}\psi_{n}(x)\,\psi_{m}(y)\,\psi_l(z)$, where the in-plane basis states are $\mathbf{1}$-D Harmonic oscillator states for $x$ and $y$ and the out-of-plane basis states are given by solutions of the infinite potential well:
\begin{eqnarray}\label{eqn:3Dbasis}
\psi_{n}(x)&=&\frac{1}{\sqrt{2^{n}\,n!}}\frac{1}{\sqrt{L_x\sqrt{\pi}}}e^{-\frac{x^2}{2L_x^2}}H_{n}\!\!\left(\frac{x}{L_x}\right),\ x\in\{-\infty,\infty\}\nonumber\\
\psi_{m}(y)&=&\frac{1}{\sqrt{2^{m}\,m!}}\frac{1}{\sqrt{L_y\sqrt{\pi}}}e^{-\frac{y^2}{2L_y^2}}H_{m}\!\!\left(\frac{y}{L_y}\right),\ y\in\{-\infty,\infty\}\nonumber\\
\psi_l(z)&=& \begin{cases}
      \cos\!\left(\frac{(l+1)\pi z}{L_z}\right) & l\ \text{is even}\\
      \sin\!\left(\frac{(l+1)\pi z}{L_z}\right) & l\ \text{is odd}\\
    \end{cases}\  ,z\in \left\{-\frac{L_z}{2},\frac{L_z}{2}\right\}
\end{eqnarray} 
The indices $n,m,l$ in Eqn.~\ref{eqn:3Dbasis} can take integer values $0,1,2,3..$ etc. The hole spinors represent the $j{=}\frac{3}{2}$ spin states: $\left|\phi_j\right\rangle\in\left\{\left|\frac{3}{2}\right\rangle,\,\left|-\frac{3}{2}\right\rangle,\,\left|\frac{1}{2}\right\rangle,\,\left|-\frac{1}{2}\right\rangle\right\}$. When operated at low in-plane $\mathbf{B}$ we are in the $\omega_c\ll\omega_0$ limit, with $\omega_c$ the cyclotron frequency, so any effect of the in-plane magnetic field on the dot size are generally irrelevant (they are taken into account in our formalism). This means the Fock-Darwin solutions have a one-on-one analogy to the 2D Harmonic oscillator solutions in Eqn.~\ref{eqn:3Dbasis}. The in-plane basis states $\Psi_n(x)\Psi_m(y)$ are ordered according to their energy $\propto(n+m+1)$. We find converging solutions to Eqn.~\ref{eq:totalQDH} by considering $55$ in-plane basis states, i.e. $(n+m)\in[0,9]$; and $15$ out-of-plane $\Psi_l(z)$ basis states, i.e. $l\in[0,14]$. We note that the $(n+m)=0$ level has no degeneracy; but considering the degeneracies of $(n+m)=1,2....9$; the simulation spans $55$ in-plane levels. The numerical diagonalisation of the resultant $3300\times3300$ Hamiltonian yields the energy levels of the hole quantum dot: $H_{QD}\left|\Psi_{QD}\right\rangle{=}\lambda_E\left|\Psi_{QD}\right\rangle$. 

In the Schrieffer-Wolff approximation, which assumes the longitudinal confinement is much stronger than the in plane confinement, theoretical models show that the in-plane magnetic field gives rise to an effective spin-orbit coupling whose magnitude is proportional to $B$.\cite{kloeffel2018direct,wang2021optimal} Recent experimental works have shown unambiguously that the $g$-factor of a 2D hole gas is a strong function of density,\cite{marcellina2018electrical,akhgar2019g}, which can be captured by effective 2D theoretical models\cite{winkler2004spin,miserev2017dimensional,miserev2017mechanisms, marcellina2017spin}. Moreover, the effective spin-orbit interaction due to the orbital magnetic field terms has a highly non-trivial interplay with the Rashba spin-orbit interaction stemming from the top-gate potential.\cite{winkler2003spin,winkler2004spin, miserev2017mechanisms} For a hole QD in an in-plane $B$, as considered here, the orbital magnetic field terms also contribute to the spin dynamics. Nevertheless, this contribution, as the following sections make clear, cannot be captured by a naive Schrieffer-Wolff transformation, because the orbital magnetic field terms couple the in-plane and out-of-plane dynamics in a way that makes them inseparable: if one first reduces the 3D Hamiltonian to an effective $2 \times 2$ Hamiltonian for a 2D hole gas, and then attempts to understand QD dynamics based on this effective 2D Hamiltonian (in analogy with electron systems) all the physics of the orbital magnetic field terms is lost. Hence a full 3D theoretical model is essential to understand the full spin dynamics of a hole quantum dot in an in-plane magnetic field. The model we present in this work can treat arbitrary quantum dot sizes, magnetic field strengths and orientations.

We comment briefly on the choice of spatial Ge basis functions. Variational analyses of the $z$-wave function incorporating the top gate potential have successfully described Ge hole QDs in Refs.~\cite{marcellina2017spin, wang2021optimal}, but the variational model is hard to extend to a full 3D numerical analysis due to the complicated form of the variational excited states. For example, the Airy function\cite{wang2022modelling,li2020hole} provides the exact solution if the $z$ confinement is modelled as a triangular potential well but can yield a residual Rashba spin-orbit interaction at non-zero top gate potential ($F_z\!>\!0$), requiring a careful choice and implementation of boundary conditions. In the present paper we describe the $z$ confinement using an infinite square well augmented by a linear electrostatic potential that accounts for the top gate, and consider top gate fields up to 50MV/m (although values up to 100 MV/m can also be studied with this method). Ref.~\onlinecite{wang2022modelling} used a sophisticated model that incorporates Fowler-Nordheim tunnelling, which is beyond the scope of the present study, as explained below. We stress that the range of $F_z$ is at the lower end of what we consider here (up to 2.5 MV/m), and thus our studies can be regarded as complementary. 
% (NOTE FOR DIMI - SEE ALSO COMMENTS: 1Mv/m ~ 1E11/CM2, SO 50 ~ 5e12. THIS IS IN PRINCIPLE OK, BUT I SUSPECT FOWLER NORDHEIM TUNNELLING WILL BECOME AN ISSUE WELL BEFORE THEN)

\section{Circular Quantum dot}
\paragraph*{\textcolor{teal!70!black}{Qubit Zeeman splitting.}}\label{sec:Zeeman}
We solve the full 3D Hamiltonian in an external magnetic field along $x$ for a QD with following dimensions: $L_x{=}50\,\text{nm}$,$L_y{=}50\,\text{nm}$,$L_z{=}11\,\text{nm}$. 
\begin{figure}[tbp]
% \subfloat[]{\begin{minipage}[c]{\linewidth}
% \includegraphics[width=3.5 in, height=2.1 in]{Larmor3Dsci.png}
% \label{fig:lf_3d}
% \end{minipage}}
% \\[-.5 cm]
\subfloat[]{\begin{minipage}[c]{.52\linewidth}
\includegraphics[width=1.58 in, height= 2.05 in]{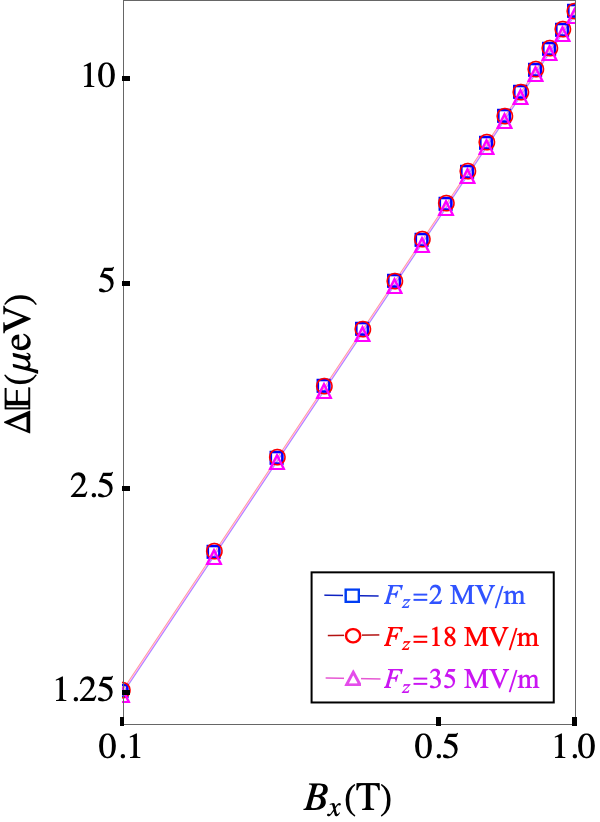}
\label{fig:ss_ip_lowB}
\end{minipage}}
\hfill
\subfloat[]{\begin{minipage}[c]{.48\linewidth}
\includegraphics[width=1.61 in, height= 2.05 in]{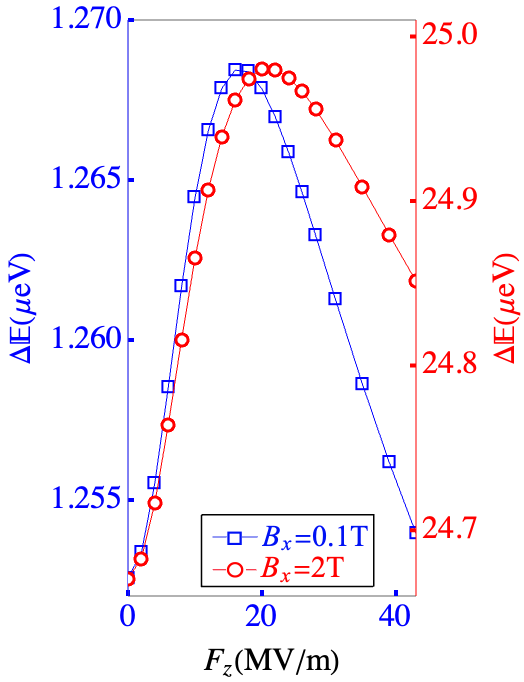}
\label{fig:ss_ip_highB}
\end{minipage}}
\caption{
%a)Variation of qubit Larmor frequency $\Delta\mathbb{E}$=$\mathbb{E_1}-\mathbb{E_0}$ ($\mu\text{eV}$) with top-gate voltage $F_z$ (MV/m) and applied in-plane magnetic field $B_x$ (T). 
\textbf{Qubit Zeeman splitting of hole spin-qubit in $B_\parallel$.} a) Variation in the qubit Larmor frequency $\Delta\mathbb{E}$ ($\mu\text{eV}$) with the in-plane magnetic field $B_x$(T). The Zeeman splitting shows a monotonic linear trend with $B_x$ at $F_z{=}2\,\text{MV/m}$, $F_z{=}18\,\text{MV/m}$ and $F_z{=}35\,\text{MV/m}$ top-gate potentials respectively; with in-plane $g_\parallel{\sim}\,0.22$. No significant difference occurs in the qubit Larmor frequency for different top gate field. b) The in-plane $g$-factor exhibits an extremum as a function of the top gate at $F_z{=}18\,\text{MV/m}$ and $F_z{=}21\,\text{MV/m}$, respectively for $B_x{=}0.1\text{T}$ and $B_x{=}2\text{T}$.}
\label{FIG:Zeeman}
\end{figure}
The ground state is given by $\left|\Psi_{QD}\right\rangle=\sum_i c_i\psi_i(x,y,z)\left|\phi_j\right\rangle$. We evaluate the QD ground state and first excited state, labelling them as state $\left|\mathbb{0}\right\rangle$ with eigenenergy $\mathbb{E_0}$ and $\left|\mathbb{1}\right\rangle$ with eigenenergy $\mathbb{E_1}$ repectively. The ground state and the first excited state are of heavy hole-type with admixtures due to heavy hole-light hole coupling, and split by an amount $\Delta\mathbb{E}{=}\mathbb{E_1}-\mathbb{E_0}$, which we refer to as the qubit Larmor frequency, while continuing to express it in units of energy. The HH-LH splitting $\Delta_{LH}\sim 60 \text{meV}$ is governed by the $z$-energy bands in the quasi 2D dot limit % (HOW BIG IS THIS ROUGHLY? I'D LIKE TO SAY THAT LH-HH SPLITTING IS ENHANCED DUE TO STRAIN AND IS MUCH LARGER THAN THE QUANTUM DOT LEVEL SPACING, SO THAT...)
%added the typical HH-LH splitting for Ge -- Abhik.
, so that many in-plane (quantum  dot) levels are contained between any two out-of-plane (quantum well) levels. This allows us to write $\left|\mathbb{0}\right\rangle{=}(c_1\left|0,0,0,\frac{3}{2}\right\rangle+\text{admixtures})$ and $\left|\mathbb{1}\right\rangle{=}(c_1'\left|0,0,0,-\frac{3}{2}\right\rangle+\text{admixtures})$, where the four indices denote $n,\,m,\,l,\,J_z$. When plotted against the top-gate voltage, the qubit Larmor frequency exhibits an extremum as a function of $F_z$.\cite{wang2021optimal} The in-plane magnetic field gives rise to a finite Zeeman splitting, which shows a linear trend with $B_x$ (figure~\ref{FIG:Zeeman}a).
The extrema in the qubit Larmor frequency as a function of $F_z$ in an in-plane $\mathbf{B}$ are explained by the same mechanism as for out-of-plane magnetic field operation (Figure~\ref{FIG:Zeeman}b). At small values of $F_z$ the matrix elements connecting the HH and LH states, which give rise to Rashba spin-orbit coupling, increase linearly with the gate field, while the change in the HH-LH splitting is negligible. At large values of $F_z$ the increase in the HH-LH splitting outweighs all other effects and the Rashba spin-orbit coupling decreases as a function of $F_z$. These competing effects give rise to an extremum in the qubit Larmor frequency at a certain value of the top gate electric field, where the qubit is insensitive to $z$-electric field fluctuations. The qubit Zeeman splitting is linear in $\mathbf{B}$, $\Delta\mathbb{E}{=}g_\parallel\mu_BB_x$, where the effective in-plane $g$-factor ranges between $0.215-0.219$, expected to be much smaller than out-of-plane $g$-factor; $g_\parallel\ll g_\perp$.

\paragraph*{\textcolor{teal!70!black}{EDSR.}}\label{sec:EDSR}
An alternating electric field $\tilde{\mathbf{E}}(t)$ can induce spin-flip transition between the qubit states $\left|\mathbb{0}\right\rangle$ and $\left|\mathbb{1}\right\rangle$ via electron dipole spin resonance (EDSR) when the frequency of the ac electric field matches the Zeeman splitting of the hole spin qubit, $\Delta\mathbb{E}{=}h\nu$. 
% (IS IT CORRECT THAT OUT OF PLANE ALTERNATING FIELDS CANNOT DO THIS?) It may be able to do so. I have changed the text to say an ac field induces transitions and here we focus on in-plane fields. 
% Abhik - for round 2, we could add E_z(t). 
 \begin{figure}[t]
\subfloat[]{\begin{minipage}[c]{\linewidth}
\includegraphics[width=3.2 in, height=2.3 in]{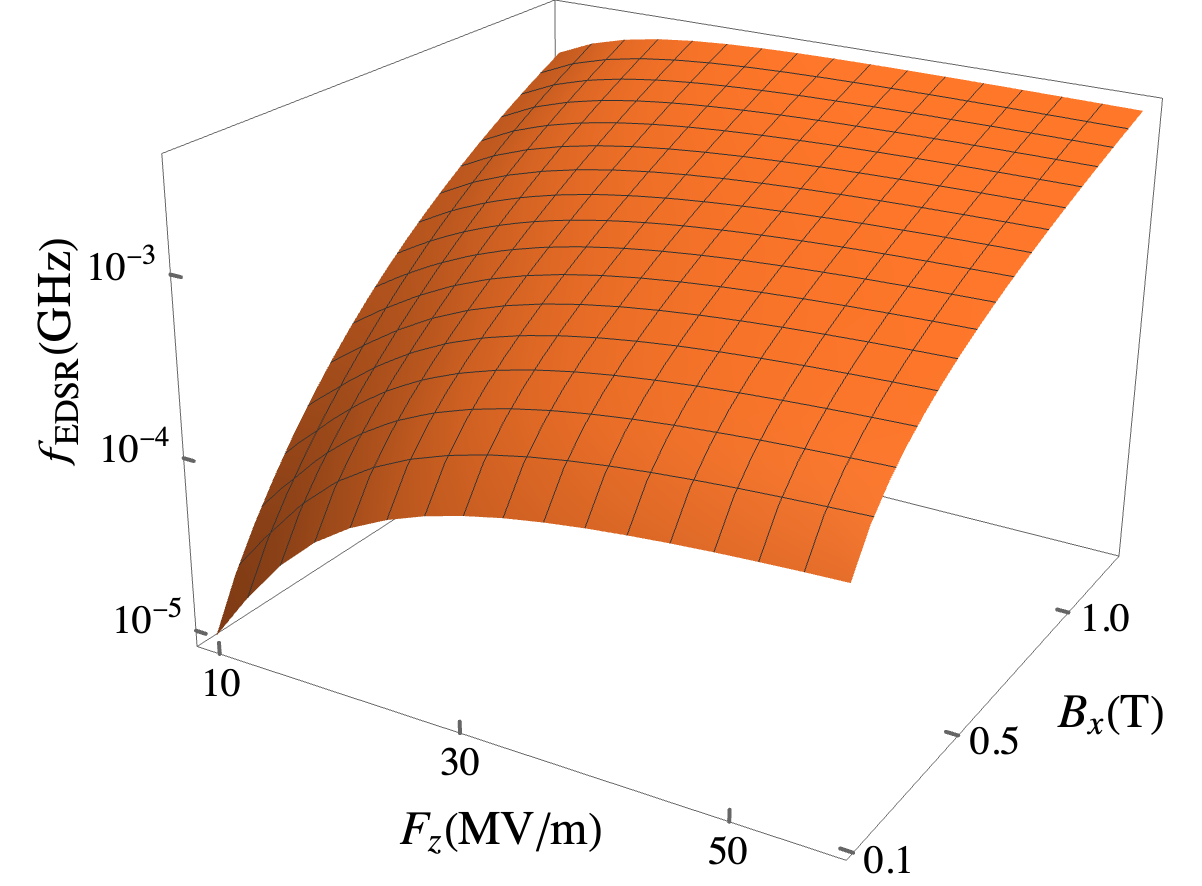}
\label{fig:EDSRrate3d}5
\end{minipage}}
\\[-.1 cm]
\subfloat[]{\begin{minipage}[c]{.52\linewidth}
\includegraphics[width=1.65 in, height= 2.2 in]{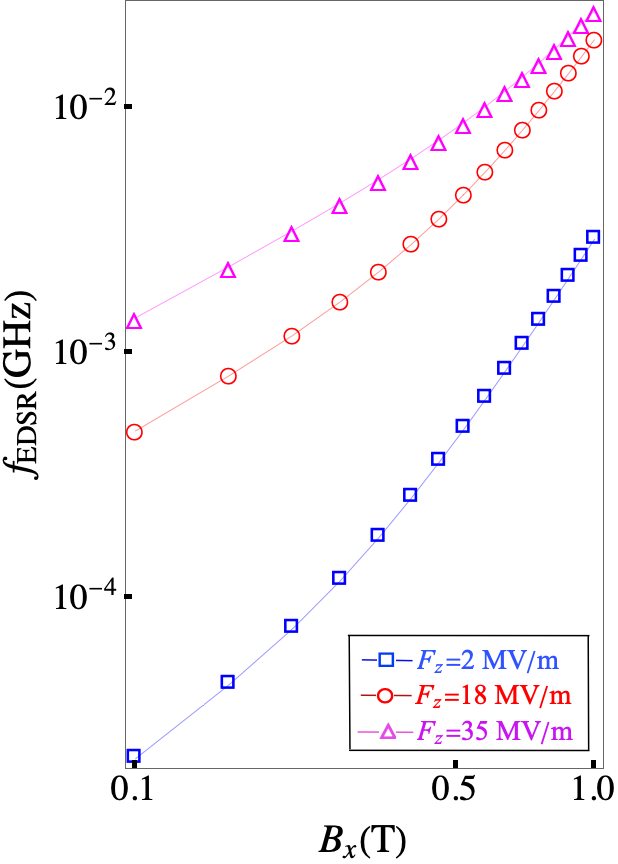}
\label{fig:EDSRFzdep_lowB}
\end{minipage}}
\hfill
\subfloat[]{\begin{minipage}[c]{.48\linewidth}
\includegraphics[width=1.6 in, height= 2.2 in]{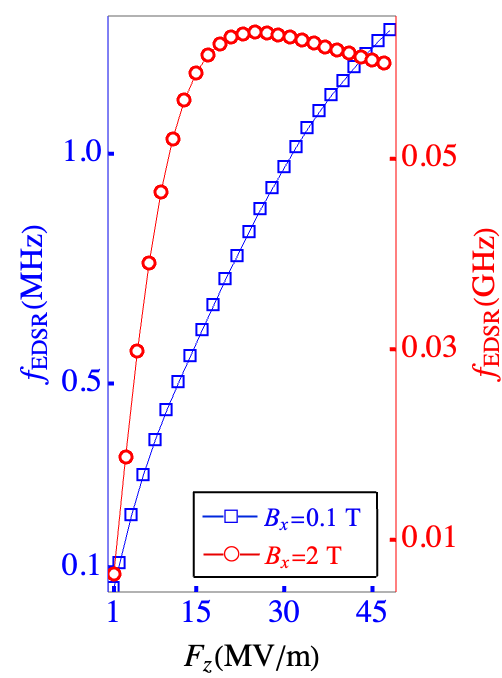}
\label{fig:EDSRFzdep_highB}
\end{minipage}}
\caption{\textbf{Electric dipole spin resonance of hole qubit in $B_\parallel.$} a) Dependence of the EDSR Rabi frequency $f_{\text{EDSR}}$(GHz) on the top gate field $F_z$(MV/m) and in-plane magnetic field $B_x$(T). The strength of the in-plane EDSR driving ac electric field is $E_{0}=10\,\text{kV/m}$. 
%have put this in the main text--Abhik.
% PHYSICS QUESTION - IF THE ZEEMAN SPLITTING VARIES WITH FIELD ORIENTATION, DOT SHAPE, APPLIED FIELD MAGNITUDE, WILL THAT NOT ALSO AFFECT THE RABI FREQUENCY?) 
% DC Yes it can affect it. 
The $z$-axis representing $f_{\text{EDSR}}$ in the $3$D plot is logarithmic. 
b) Variation of $f_{\text{EDSR}}$ with $B_x$ showing a non-monotonic dependence on the applied $B$-field as: $f_{\text{EDSR}}{=}a_fB_x{+}b_fB_x^2{+}c_fB_x^3$. The points on the plot signify results from the numerical calculation while the fitting is denoted by the dashed lines. At $F_z{=}2\,\text{MV/m}$ the fitting parameters are $a_f=0.0002,\, b_f=0.00002,\, c_f=0.003$; at $F_z{=}18\,\text{MV/m}$ they are $a_f=0.005,\,b_f=0.00006,\,c_f=0.01$ and at $F_z{=}35\,\text{MV/m}$ top-gate potential the fitting parameters are $a_f=0.01,\,b_f=0.00004,\,c_f=0.01$.
%THESE FREQUENCIES SEEM LOW.-Alex. 
c) The Rabi frequency exhibits maxima as a function of the top gate at $F_z{=}18\,\text{MV/m}$ only for higher magnetic field ($B_x{=}2\text{T}$), over the range of top-gate potentials used in this study.}
\label{FIG:EDSR}
\end{figure}
The EDSR Rabi frequency is calculated as the matrix element of the ac field between the qubit ground state (GS) and excited state (ES):
\begin{equation}
    f_{\text{EDSR}} = \left\langle\mathbb{0}\left|e\tilde{\mathbf{E}}(t)\cdot \mathbf{r}\right|\mathbb{1}\right\rangle
\end{equation}
Here we will focus on the scenario in which the alternating electric field is in the plane. In the 3D model, the effective interaction between the qubit GS and ES wavefunctions lies in the LH admixture to $\left|\pm\frac{3}{2}\right\rangle$ HH states. The admixture is the results of the two primary spin-orbit interactions in the system: structure inversion asymmetry (SIA) due to the top-gate potential $F_z$ gives rise to the first Rashba term, which stems primarily from HH-LH coupling. The second contribution to the spin-orbit interaction comes from the orbital terms due to the in-plane magnetic field $\mathbf{B}$. As $\mathbf{B}$ is ramped up, the magnitude of this latter contribution increases. For an applied oscillatory electric field of strength $E_0{=}10 \text{kV/m}$, figure \ref{FIG:EDSR}a presents the spin-flip Rabi frequency variation with the top gate field and applied $B_x$, with the key features of the EDSR frequency exhibiting a maximum at a certain value of $F_z$, as well as a nonlinear dependence of $f_{\text{EDSR}}$ on $B_x$. In figure~\ref{FIG:EDSR}b, $f_{\text{EDSR}}$ shows a non-linear variation as a function of the in-plane magnetic field. The best fit corresponds to $f_{\text{EDSR}}{=}a_fB_x{+}b_fB_x^2{+}c_fB_x^3$ at a constant top gate field. At $B_x{=}0.1\,\text{T}$ the Rabi frequency increases slowly with the top gate field. On the other hand at ($B_x{=}2\,\text{T}$) the Rabi frequency increases more rapidly with $F_z$, and the maximum shifts towards lower values of $F_z$ (Fig.~\ref{FIG:EDSR}c). We find the EDSR rate is a maximum when the electric field is parallel to the magnetic field, and vanishes when the two are perpendicular.
\begin{figure}[tbp]
%\subfloat[]{\begin{minipage}[c]{\linewidth}
%\includegraphics[width=3 in, height= 2.25 in]{RelaxationTimeCircular3D_new_new.png}
%\label{fig:T1BFzdepend}
%\end{minipage}}
%\\[-.2 cm]
\subfloat[]{\begin{minipage}[c]{.48\linewidth}
\includegraphics[width=1.65 in, height= 2.15 in]{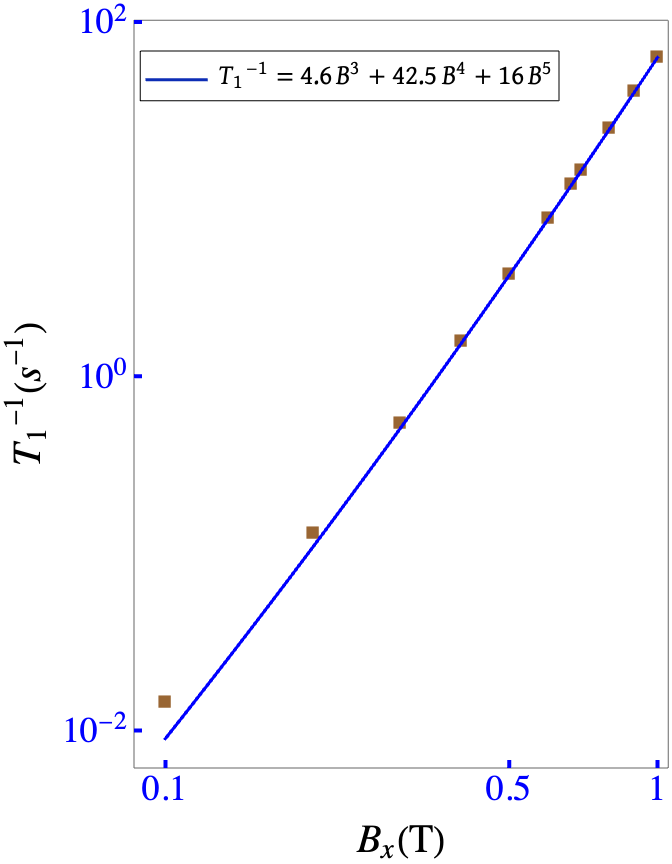}
\label{fig:T1Fzdepend}
\end{minipage}}
\hfill
\subfloat[]{\begin{minipage}[c]{.52\linewidth}
\includegraphics[width=1.65 in, height= 2.15 in]{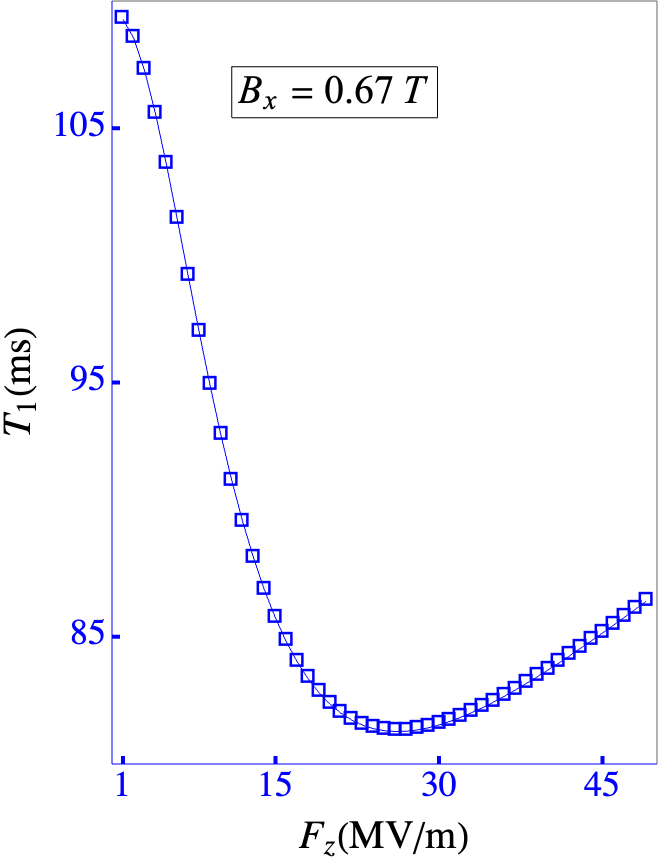}
\label{fig:T1Bdepend}
\end{minipage}}
\caption{\textbf{Phonon induced hole spin-qubit relaxation for $B_\parallel$ operation.} a) Relaxation rate $T_1^{-1}\,(\text{s}^{-1})$ vs. applied magnetic field $B_x$ (T) at the top-gate field of  $F_z{=}18\,\text{MV/m}$. The B-dependence of the relaxation rate has been fitted using the fitting parameters shown in the top inbox. b) The relaxation rate $T_1^{-1}$ vs. the top gate field $F_z$ (MV/m) for $B_x{=}0.67\,\text{T}$.}
\label{FIG:Relaxation}
\end{figure}

 The nonlinearity in $F_z$ in figure~\ref{FIG:EDSR}c reflects the quadrupole Rashba spin-orbit coupling unique to spin-3/2 systems,\cite{winkler2004spin} having no counterpart in electron spin qubits. Hole qubits in Ge quantum dots can exhibit fast EDSR of the order of $\sim\!100\,\text{ns}$, despite the fact that the in-plane $g$-factor is two orders of magnitude smaller than the out-of-plane $g$-factor. The linear term in $B_x$ in the Rabi frequency (figure~\ref{FIG:EDSR}b) stems from the Zeeman terms cubic in angular momentum $\propto q$ in Eq.~\ref{eq:Zeeman}. To see this one can write an effective $2{\times}2$ qubit Hamiltonian using the Schrieffer-Wolff (SW) transformation up to third order in small quantities:
 \begin{equation}\label{eq:effective2DEDSR}
H_{\text{eff}}^{2{\times}2}=H_0+H_{SO}+g_\parallel\mu_BB_x^3\sigma_x+V(x,y)+eE_x(t)x
 \end{equation}
 Here $H_0\!(\mathbf{k}{\rightarrow}(\mathbf{k}{+}e\mathbf{A}/\hbar)){=}[\mathcal{A}k^2\!-\!\mathcal{B}k^4\!-\!D(k_+^2\!-\!k_-^2)^2]\!I_{2{\times}2}$ is the kinetic energy including a correction $\propto k^4$,\cite{winkler2004spin} and the $D$ term represents warping of the energy contours.\cite{marcellina2017spin} The spin-orbit Hamiltonian $H_{SO}$ comprises two important $k$-cubic Rashba terms induced by the top gate field, a spherical term $\propto \alpha_{R2}$ and a cubic-symmetric correction $\propto \alpha_{R3}$. The next three terms represent the Zeeman splitting, confinement energies and the ac electric potential along the $x$-direction, respectively. Starting from this effective Hamiltonian and projecting it onto the in-plane states, one may apply the SW approximation again to obtain an effective $2 \times 2$ qubit Hamiltonian, in which the EDSR driving term appears on the off-diagonal. Reading off the EDSR Rabi frequency we find $f_{\text{EDSR}}{\propto}B_x$ to leading order, while the next order in the expansion gives $f_{\text{EDSR}} \propto B_x^3$ (Appendix.~\ref{appen:EDSR}).
 
 The significant quadratic $B_x^2$ dependence of the EDSR Rabi frequency at large $F_z$ (Fig.~\ref{fig:EDSRFzdep_highB}) is a signature of the orbital magnetic field terms. One can determine possible \textit{paths} connecting the lowest energy $\left|\mathbb{0}\right\rangle{\leftrightarrow}\left|\mathbb{1}\right\rangle$ HH-qubit states, mainly composed of $\left|0,0,0,\frac{3}{2}\right\rangle$ and $\left|0,0,0,-\frac{3}{2}\right\rangle$ with admixtures, respectively. Choosing the symmetric gauge for $\mathbf{B}{=}(B_x,0,0)$, the magnetic field induces the following transitions in the Luttinger-Kohn picture: $\left\{\left|\frac{3}{2}\right\rangle\!{\xrightarrow{B_x}}\!\left|\frac{1}{2}\right\rangle,\,\left|-\frac{1}{2}\right\rangle\!{\xrightarrow{B_x}}\!\left|-\frac{3}{2}\right\rangle\right\}$. The gate potential gives $\left\{n_z{\xrightarrow{eF_zz}}n_z{\pm}1\right\}$ which is spin conserving. The Luttinger terms $L(\mathbf{k}),\,M(\mathbf{k})$ couple HH and LH states e.g. $\left\{\left|n_x,n_y,n_z,\frac{3}{2}\right\rangle\!{\xrightarrow[L(\mathbf{k})]{k_xk_z}}\!\left|n_x\pm1,n_y,n_z\pm1,\frac{3}{2}\right\rangle\right\}$. We consider the complete SO picture in the Ge hole QD following eqn.~\ref{eq:totalQDH} to sketch a few example \textit{paths} as follows:
\begin{small}
\begin{itemize}
    \item $\left|0,0,0,\frac{3}{2}\right\rangle\!{\xrightarrow[L]{B_x}}\!\left|0,0,0,\frac{1}{2}\right\rangle\!{\xrightarrow[L]{k_yk_z}}\!\left|0,1,1,\frac{3}{2}\right\rangle\!{\xrightarrow[M]{k_xk_y}}$\\~$\left|1,2,1,-\frac{1}{2}\right\rangle\!{\xrightarrow{eF_zz}}\!\left|1,2,0,-\frac{1}{2}\right\rangle\!{\xrightarrow{eE_xx}}\!\left|0,2,0,-\frac{1}{2}\right\rangle\!{\xrightarrow[J\rightarrow J]{k_y^2}}$\\~$\left|0,0,0,-\frac{1}{2}\right\rangle\!{\xrightarrow[L]{B_x}}\!\left|0,0,0,-\frac{3}{2}\right\rangle$.

    \item $\left|0,0,0,\frac{3}{2}\right\rangle\!{\xrightarrow[L]{k_yk_z}}\!\left|0,1,1,\frac{1}{2}\right\rangle\!{\xrightarrow{eE_xx}}\!\left|1,1,1,\frac{1}{2}\right\rangle\!{\xrightarrow[J\rightarrow J]{B_xyk_z}}$\\~$\left|1,0,2,\frac{1}{2}\right\rangle\!{\xrightarrow{(eF_zz)^2}}\!\left|1,0,0,\frac{1}{2}\right\rangle\!{\xrightarrow[M]{k_xk_y}}\!\left|0,1,0,-\frac{3}{2}\right\rangle\!{\xrightarrow[J\rightarrow J]{B_xyk_z}}$\\~$\left|0,0,1,-\frac{3}{2}\right\rangle\!{\xrightarrow{eF_zz}}\!\left|0,0,0,-\frac{3}{2}\right\rangle$.
\end{itemize}
\end{small}
These \textit{paths} explain the quadratic $B_x^2$ dependence at large $F_z$ in the presence of the ac electric field. 
%why? 
% the standard rashba is explained by \alpha E.(\sigma x k), \sigma and k has the same symmetry?(ref. winkler) and k x k = B.
%strain and EDSR
%EDSR usually due to in-plane electric field, $E_z(t)$ no EDSR at sweet spot.

\paragraph*{\textcolor{teal!70!black}{Phonon mediated Relaxation.}}\label{sec:relaxation}

The relaxation mechanism in Ge hole QD qubits is well explained by acoustic phonon coupling to the hole spins through the valence band deformation potential $\mathcal{D}_{i,j}$ of Ge. There are no piezo-electric phonons in Ge, but the hole spin interacts with the thermal bath of the bulk phonons via the hole-phonon Hamiltonian $H_{h-ph}$, where $\alpha\in\{l,t,w\}$ denotes the polarization directions of the phonon and $\mathbf{q}$ is the phonon wave vector. We write $H_{h-ph}{=}\sum_{i,j} \mathcal{D}_{i,j} \varepsilon^\alpha_{i,j}(\bm{r})$ with $\varepsilon^\alpha_{i,j}(\bm{r})= \sum_q\frac{i}{2} \sqrt{\frac{\hbar}{2\rho NV\omega_{\mathbf{q},\alpha}}} \left(q_i\hat{c}_j+q_j\hat{c}_i\right) e^{i \bm{q} \cdot \bm{r}}\sqrt{N_q^\alpha+1}$. The relaxation rate $1/T_1$ is written as:
\begin{equation}
    \frac{1}{T_1}=\frac{2\pi}{\hbar}\sum_{\alpha,\mathbf{q}}\left|\left\langle\mathbb{0}\right|H_{h-ph}\left|\mathbb{1}\right\rangle_\alpha\right|^2\delta\left(\Delta\mathbb{E}-\hbar\omega_{\alpha,\mathbf{q}}\right)
\end{equation} 
\begin{figure}[tbp]
\subfloat[]{\begin{minipage}[c]{0.5\linewidth}
\includegraphics[width=1.57 in, height= 2.2 in]{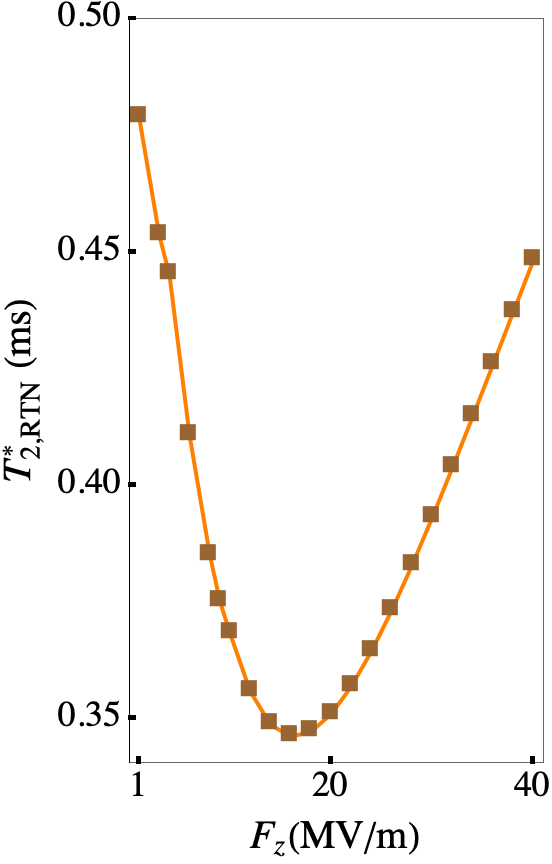}
\end{minipage}}
\hfill
\subfloat[]{\begin{minipage}[c]{0.5\linewidth}
\includegraphics[width=1.65 in, height= 2.2 in]{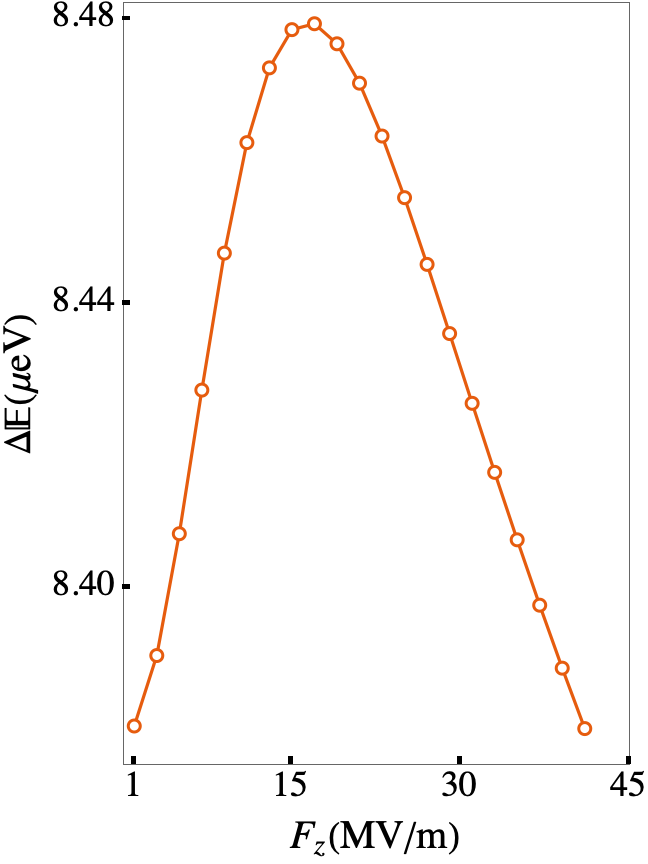}
\end{minipage}}
\caption{\textbf{Qubit dephasing in an in-plane magnetic field:} a) Variation of qubit dephasing time $T_{2,\text{RTN}}^*$ (ms) due to random telegraph noise (RTN) from a nearby in-plane single charge defect 200 nm away from the qubit w.r.t. top gate field $F_z$ (MV/m). b) Qubit Zeeman splitting $\Delta E\,(\mu\text{eV})$ vs. applied top-gate field $F_z$ (MV/m). The applied magnetic field is in the plane with the magnitude $B_x{=}0.67\,\text{T}$.
% CAN THE FIGURE START AT 1 NOT 5? 
% Abhik is working on it
%An in-plane magnetic field of $B_x{=}0.7\,\text{T}$ is used. 
% WHY IS THE IN-PLANE FIELD DIFFERENT FOR EVERY FIGURE? 
% Let's use 0.67T here as well. 
The dephasing time reaches a minimum at $F_z{=}18\,\text{MV/m}$. 
% BUT KEY POINT - IT HARDLY VARIES!! Fz VARIES BY FACTOR 6, RELAXATION TIME BY 5 PERCENT. WHAT IS SIGNIFICANCE OF THIS? THIS POINT NOT DISCUSSED IN TEXT.
% It does vary by 50%. 
}
\label{FIG:RTNDephasing}
\end{figure}
We calculate the relaxation rate using our semi-analytical 3D QD model by computing in real space the overlap integral $\left\langle\mathbb{0}\right|H_{h-ph}\left|\mathbb{1}\right\rangle_\alpha$ due to position-dependent local strain, followed by the scattering integral in the phonon wave vector $q$-space for a specific polarization direction. The full analytical integrations and dipole approximation calculations are detailed in Appendix.~\ref{appen:relaxation}. Figure~\ref{FIG:Relaxation}a shows the nonlinear variation of the relaxation rate $T_1^{-1}$ w.r.t. the external magnetic field $B_x$, with $B^3$, $B^4$ and $B^5$ terms present in the fitting obtained from the full 3D numerical model. While the $B^3$ and $B^5$ dependence are explained from the first two terms in the dipole approximation, the $B^4$ term is understood from the orbital $B$-admixture. A minimum relaxation time $T_1\sim 80\,\text{ms}$ at an in-plane magnetic field $B_x{=}0.67\,\text{T}$ is obtained. This result from the theory compares well with a single hole relaxation time measurement in Ge of over 30 ms and a five-hole relaxation time of approximately 1 ms by Lawrie et al.\cite{lawrie2020spin} The magnetic field in the experimental setup\cite{hendrickx2020single,lawrie2020spin} is $B{=}0.67\,\text{T}$, as also used in Fig.~\ref{FIG:Relaxation}a. There exists a minimum in $T_1$ in the range of $F_z$ considered here for $B_x{=}0.67\,\text{T}$ (figure~\ref{FIG:Relaxation}b). At this minimum the Rabi ratio $T_1/T_\pi {\approx} 2\times 10^5 $, where $T_\pi$ is the time required for an EDSR $\pi$-rotation, demonstrating that fast Rabi oscillations can be achieved without sacrificing $T_1$. 
\begin{figure}[tbp]
\subfloat[]{\begin{minipage}[c]{0.5\linewidth}
\includegraphics[width=1.55 in, height= 2.3 in]{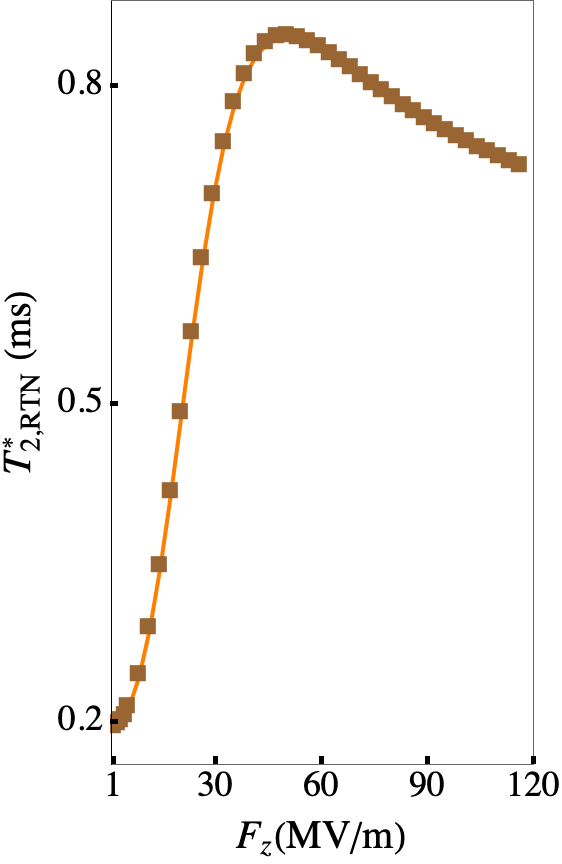}
\end{minipage}}
\hfill
\subfloat[]{\begin{minipage}[c]{0.5\linewidth}
\includegraphics[width=1.55 in, height= 2.3 in]{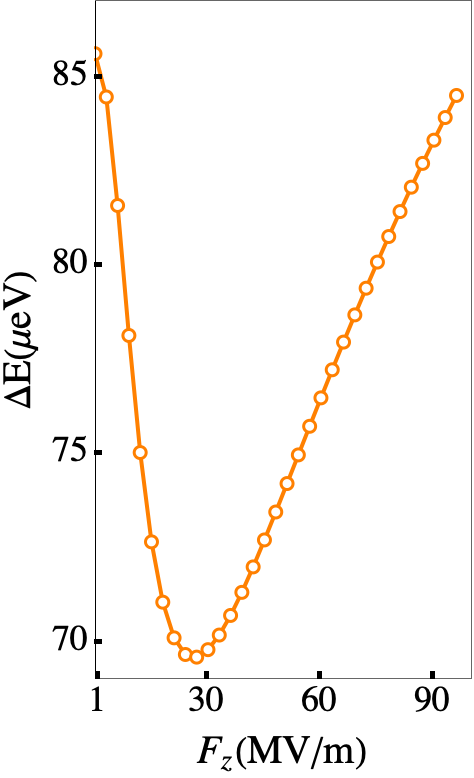}
\end{minipage}}
\caption{\textbf{Qubit dephasing in an out-of-plane magnetic field:} a) Variation of qubit dephasing time $T_{2,\text{RTN}}^*$ (ms) due to random telegraph noise (RTN) from a nearby in-plane single charge defect 200 nm away from the qubit w.r.t. top gate field $F_z$ (MV/m). b) Qubit Zeeman splitting $\Delta E\,(\mu\text{eV})$ vs. applied top-gate field $F_z$ (MV/m). The applied magnetic field is in the plane with the magnitude $B_x{=}0.1\,\text{T}$. In this case, the dephasing time exhibits a maximum in top-gate field.  
}
\label{FIG:RTNDephasingOP}
\end{figure}
\paragraph*{\textcolor{teal!70!black}{Random Telegraph Noise (RTN) Dephasing.}}\label{sec:dephasing}
The large spin-orbit coupling exposes the hole spin qubit to charge noise. The dephasing time $T^*_{2,\text{RTN}}$ is evaluated from the fluctuation in the qubit energy gap, denoted by $\delta\omega$, caused by the screened potential $U_s(\bm r)$ of a nearby single charge defect in the 2DHG.\cite{maman2020charge,culcer2009dephasing,ramon2022qubit,roszak2019detect,malkoc2022charge} The mathematical formulation of $U_s(\bm r)$ is given in Appendix.~\ref{append:dephasing}. The matrix elements $\left\langle n,m,l|U_{s}(\bm{r})|n',m',l'\right\rangle$ are added to the full Hamiltonian, and the $3300\times 3300$ matrix is diagonalised to evaluate the qubit energy splitting in the presence of the charge defect as $\Delta\mathbb{E}+\delta\omega$. The dephasing rate is:
\begin{equation}
    \left(T^*_{2,\text{RTN}}\right)^{-1}{=}\frac{(\delta\omega)^2\tau}{2\hbar^2}
\end{equation}
where the defect switching time is taken as $\tau{=}10^3\, t_{Rabi}$. This picture assumes the most significant contribution to RTN comes from charge defects away from the top gate, close to the qubit plane; hence we consider a single charge defect in the qubit plane situated $200$ nm away from the center of the qubit. Fluctuating single charge defects right above the qubit will be screened by the presence of the top-gate, where the image charge changes the interaction to a much weaker dipole interaction. In contrast fluctuating charges in the plane of the quantum well are less effectively screened by surface gates, and may be the dominant source of charge noise. \cite{culcer2013dephasing}

It is evident that for hole qubit in an in-plane magnetic field, the dephasing time $T_{2,RTN}^*$ actually decreases as a function of the top gate electric field $F_z$, and reaches a minimum at a certain value of this field (Fig.~\ref{FIG:RTNDephasing}a), in other words, a coherence hot spot. The location of this dephasing time hotspot is closely related to the extremum in the qubit Zeeman splitting (Fig.~\ref{FIG:RTNDephasing}b). This behaviour is in sharp contrast to hole spin qubits in a perpendicular magnetic field, where the qubit exhibits a sweet spot at a certain value of the top gate field (Fig.~\ref{FIG:RTNDephasingOP}a), at which its sensitivity to noise vanishes to leading order in the noise strength, and dephasing time $T_{2,RTN}^*$ reaches a maximum. The location of the sweet spot in $F_z$ for out-of-plane qubit operation is closely related to an extremum in the qubit Zeeman splitting (Fig.~\ref{FIG:RTNDephasingOP}b).  

\paragraph*{\textcolor{teal!70!black}{$B_\perp$ vs. $B_\parallel$ coherent qubit operation.}}\label{sec:outofplanesupremecy} In the context of qubit coherence one must distinguish between extrema in the qubit Zeeman splitting and actual sweet spots in the coherence time. It is important to recall that in a spin qubit the dephasing time $T_2^*$ depends on the magnitude of the magnetic field. 
% IF B SO IMPORTANT THEN WHY NOT SHOW EFFECTS OF FIELD MAGNITUDE (AND POSSIBLY ORIENTATION) IN FIGUYE 5?. 
% Because that will be done in the coherence paper with Sankar. 
This follows from time-reversal symmetry considerations, since the combination of charge noise and spin-orbit coupling cannot give rise to an energy difference between qubit states that form a Kramers doublet. The magnetic field dependence involves both the Zeeman terms and the orbital vector potential terms, a fact that is responsible for the main difference between in-plane and out-of-plane magnetic fields with regard to qubit dynamics: the make up of the ground and first excited states is very different when the magnetic field is in the plane and when it is out of the plane. 
% I DON'T SEE HOW THE END OF THIS PARAGRAPH IS RELATED TO THE BEGINNING...? IN FACT I DON'T UNDERSTAND THE ARGUMENT BEING MADE AT ALL. WE START BY SAYING THAT I MUST REMEMBER THAT DEPHASING IS A FUNCTION OF B. BUT DO WE MEAN MAGNITUDE OR ORIENTATION? IF THE LATTER THEN WHY TALK ABOUT KRAMERS DOUBLETS?
% I have specified now that it is a function of the magnitude of the magnetic field. It may vary with the orientation of B but that is irrelevant for the present argument. 

For an out of plane magnetic field the hole $g$-factor is large, having a textbook value of 20.4 for Ge.\cite{winkler2003spin, lu2017effective} With the magnetic field out of the plane one can understand the physics qualitatively by considering an approximate decomposition of in-plane and out-of-plane dynamics by means of a Schrieffer-Wolff transformation \cite{wang2021optimal} % BUT WE SAID THAT SW APPROACH IS INVALID? 
The picture that emerges is that the top gate electric field primarily affects spin dynamics in the plane by enabling a Rashba term. In a quantum dot this Rashba term is responsible for a renormalization of the $g$-factor. In other words, one can think of the magnetic field terms as providing the qubit Zeeman splitting, and the Rashba spin-orbit terms as renormalizing this Zeeman splitting. Background charge fluctuations generating an electric field perpendicular to the plane are the biggest danger for this qubit, because they directly affect the Rashba interaction and through it the $g$-factor, generating pure dephasing. A more detailed analysis of hole spin qubit in $B_\perp$ \cite{wang2021optimal} reveals that in-plane charge fluctuations do not produce pure dephasing to leading order. % YOU JUST SAID THE EXACT OPPOSITE ONE SENTENCE AGO! WHY PRESENT THE FORMER IF YOU IMMEDIATELY CONTRADICT IT? 
% I did not say the opposite. This is all for B-\perp, Zhanning's paper. 
Hence for $B_\perp$ operation, when the qubit Zeeman splitting is at an extremum with respect to the top gate electric field (Fig.~\ref{FIG:RTNDephasingOP}b), 
% WHICH FIGURE IN THE PRESENT PAPER SHOWS THIS? 
% It might help if we add a figure for B_\perp.
the qubit is protected against noise and one also observes a sweet spot in $T_2^*$ in the vicinity of this point (Fig.~\ref{FIG:RTNDephasingOP}a). 
% FOR ANY READER TO UNDERSTAND THIS PARAGRAPH THEY WILL NEED TO HAVE THE WANG PAPER RIGHT NEXT TO THEM.

On the other hand, for hole spin qubit in $B_\parallel$, we recall that to a first approximation the in-plane $g$-factor is zero, hence the entire qubit Zeeman splitting is given by coupling to the excited states. This coupling involves Luttinger spin-orbit terms, the orbital magnetic field terms, the top gate electric field, and any other electric fields present in the system. The orbital terms due to the magnetic field mix the in-plane and out-of-plane coordinates regardless of the gauge choice. There is no clear separation between in-plane and out-of-plane dynamics, and no suitable Schrieffer-Wolff transformation from the 3D picture to the asymptotic 2D limit. One may at best envisage a combined Rashba-Zeeman interaction with contributions from all the components of the electric field, not just the top gate. The qubit states contain a strong admixture of all the higher orbital excited states in all three directions, which exposes the qubit to all components of the electric field of the defect. Thus, even though one can still identify an extremum in the qubit Zeeman splitting as a function of the top gate field (Fig.~\ref{FIG:RTNDephasing}b), this does not offer protection against noise and does not constitute a sweet spot (Fig.~\ref{FIG:RTNDephasing}a). It only protects against noise fields perpendicular to the plane, without offering any protection against the in-plane electric field of a defect. We check explicitly that for a defect that produces only an out-of-plane electric field at the qubit location the sensitivity to this out-of-plane noise is minimised at the extremum in the Zeeman splitting. We have also checked that the qubit is not shielded from the in-plane electric field of the defect at this extremum: there is nothing special about the extremum from this perspective. We note that in an experimental sample exposed to an ensemble of defects it is possible for the net in-plane electric field to cancel out, or nearly cancel out. Hence, to achieve a more complete understanding of coherence, it is vital to consider a realistic configuration leading to $1/f$ noise. In light of this, and of additional complexities identified recently in modelling hole spin coherence,\cite{Shalak2023HoleRTN} we defer the full theory of hole spin coherence in the presence of $1/f$ noise to a future publication. 

We note that our findings appear to agree with recent experimental work reporting sweet spot operation of a Ge hole spin qubit \cite{Nico2023SweetSpot} as well as strong anisotropy in the noise sensitivity. Sensitivity to charge noise is found to increase significantly when the qubit is operated in an in-plane magnetic field. This is in agreement with the finding of the present paper that in-plane magnetic fields expose the qubit to noise much more strongly than out-of-plane magnetic fields, leading to the coherence hot spot seen in Fig.~\ref{FIG:RTNDephasing}a. Remarkably, the dominant source of noise in Ref.~\onlinecite{Nico2023SweetSpot} is believed to lie directly above the qubit, implying charge fluctuations predominantly in the perpendicular electric field component, and suggesting the qubit was not operated in the sweet spot for out-of-plane charge fluctuations. Nevertheless, a full comparison between theory and experiment is premature at this stage, given that tilting of the $g$-tensor and local strain have not been considered in the present work.

% SO IF THE PRIMARY SOURCE OF NOISE IS Fz, USE BPERP, OTHERWISE USE bPARALLEL? IS THE KEY PHYSICS THAT THE MAGNETIC FIELD SHOULD BE PARALLEL TO THE ELECTRIC FIELD NOISE? IF SO, A VECTOR MAGNET COULD BE USED TO DISTINGUISH BETWEEN DIFFERENT SOURCES OF CHARGE NOISE...

%\section{Entanglement}
%Implications for entanglement via dipole-dipole and CQED.
\begin{figure}[tbp]
\subfloat[]{\begin{minipage}[c]{\linewidth}
    \includegraphics[width=3 in, height=1.9 in]{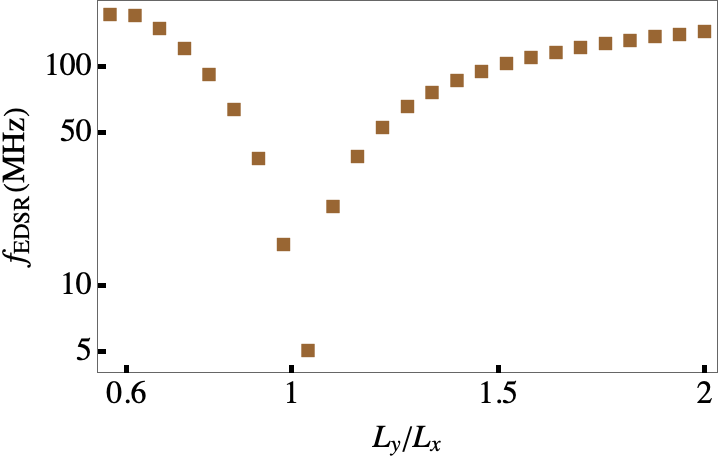}
\label{fig:ellipt_aspect}
\end{minipage}}
\\[-.3 cm]
\subfloat[]{\begin{minipage}[c]{.5\linewidth}
\includegraphics[width=1.56 in, height= 2.08 in]{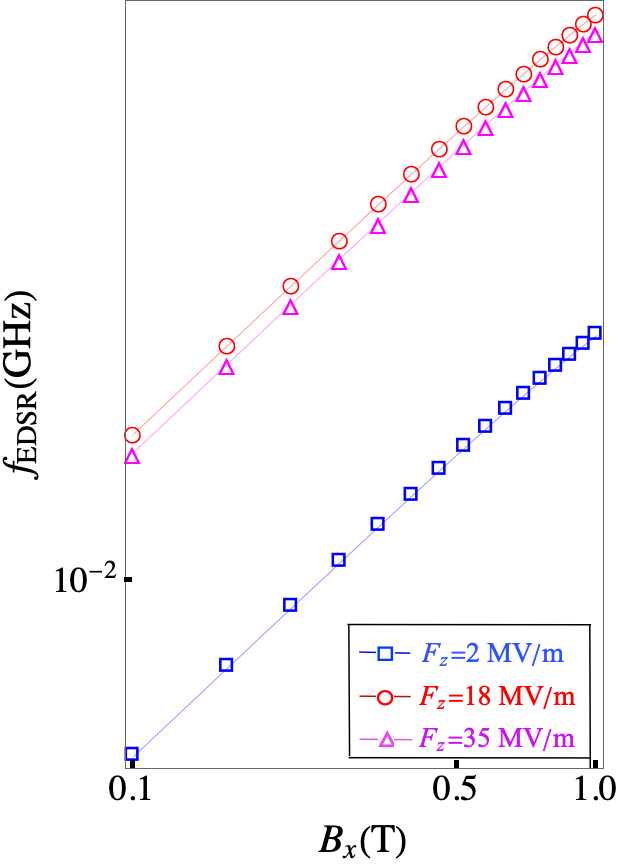}
\label{fig:ellipt_b}
\end{minipage}}
\hfill
\subfloat[]{\begin{minipage}[c]{.5\linewidth}
\includegraphics[width=1.62 in, height= 2.1 in]{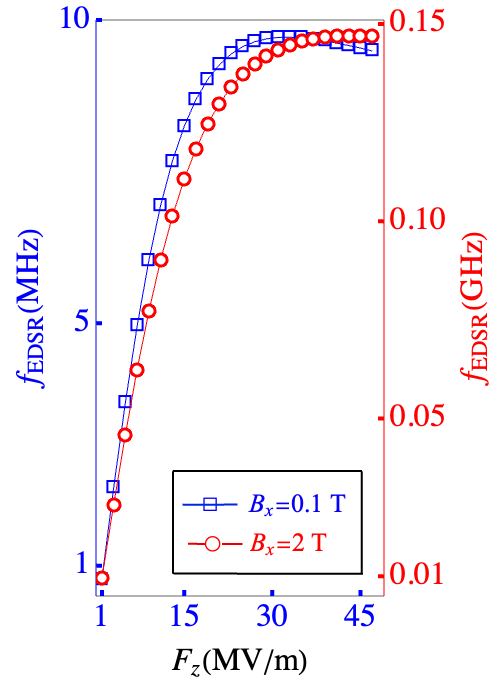}
\label{fig:ellipt_fz}
\end{minipage}}
\caption{a)Variation of the EDSR Rabi frequency $f_{\text{EDSR}}$ with the aspect ratio $L_y/L_x$ of the hole QD. The confinement in x is fixed at $L_x{=}50\text{nm}$ while $L_y$ is varied. The calculation is done for $B_x{=}0.7\text{T}$. The EDSR driving field is applied along the $\hat{x}$-direction. $f_{\text{EDSR}}$ is a minimum when the dot is circular, i.e. $L_x{=}L_y{=}50 \text{nm}$, while elliptical dots show an improvement of $>\!1$ order of magnitude in the EDSR spin-flip rate at high ellipticities. % IS THIS RESULT REALLY TRUE FOR A 5CM BY 5CM QUANTUM DOT? 
% Abhik - made some changes here, hopefully it sounds coherent.
% Dimi - I don't understand the 5cm comment. 
b) Linear monotonic dependence of $f_{\text{EDSR}}$ on the applied in-plane magnetic field $B_x$ for an elliptical dot with $L_y/L_x{=}2,\,L_x{=}50 \text{nm}$ for three different values of the top gate potential: $F_z{=}2\,\text{MV/m}$, $F_z{=}18\,\text{MV/m}$ and $F_z{=}35\,\text{MV/m}$. c) Variation in $f_{\text{EDSR}}$ with the top gate electric field for an elliptical dot at low ($0.1\,\text{T}$) and high ($2\,\text{T}$) in-plane magnetic fields, displaying a maximum EDSR Rabi frequency at $F_z{=}18\,\text{MV/m}$ and $F_z{=}21\,\text{MV/m}$, respectively.} 
% IS IT POSSIBLE TO START THE FIGURE AT 1MV/M? Abhik is working on this.
\label{FIG:Ellipt}
\end{figure}

\section{Elliptical Quantum dot}\label{sec:ellipt}
Introducing asymmetry into the planar confinement, i.e. having one lateral confinement potential stronger than the other will bring in additional sources of structure inversion asymmetry (SIA). For such elliptical hole QDs, the resultant Rashba spin-orbit interaction is stronger, bridging the gap between planar QD and nanowires \cite{bosco2021squeezed} in terms of fast gate operations. A theoretical understanding of QD ellipticity for holes is thus important. Insight into our numerical results can be obtained from the effective $2{\times}2$ spin-orbit Hamiltonian in Eqn.~\ref{eq:effective2DEDSR}, which we rewrite as:
\begin{eqnarray}\label{eq:effectiveSO}
    H_{SO}&{=}&\alpha_{R2}\!\left[k_+^3\sigma_-{-}k_-^3\sigma_+\right]\!+\!\alpha_{R3}\!\left[\{k_+^2,k_-\}\sigma_+{-}\{k_+,k_-^2\}\sigma_-\right]\nonumber\\
    &&\text{where,}\ \,\,\  \alpha_{R2}{=}\frac{e\hbar^4}
    {m_0^2}\gamma_3\overline{\gamma}S_{hh};\    \,
    \alpha_{R3}{=}\frac{e\hbar^4}{m_0^2}\gamma_3\delta S_{hh};
\end{eqnarray}
with the spherical Luttinger parameter $\overline{\gamma}{=}(\gamma_2+\gamma_3)/2$, the cubic-symmetry parameter $\overline{\delta}{=}(\gamma_3-\gamma_2)/2$, and $S_{hh}$ the sub-band interaction in $3^\text{rd}$ order perturbation theory (Appendix.~\ref{appen:EDSR}). In a circular dot the cubic-symmetry correction $\alpha_{R3}$ is responsible for EDSR, while the spherical term $\alpha_{R2}$ does not contribute. In contrast, in an elliptical dot the $\alpha_{R2}$ term is nonzero. From Eqn.~\ref{eq:effectiveSO}, $\alpha_{R3}=(\delta/\overline{\gamma})\,\alpha_{R2}$; using the lattice parameters of Ge we evaluate $\alpha_2{\approx}10\,\alpha_3$. We present the results for a dot size of $L_z{=}11\,\text{nm}$, $L_x{=}50\,\text{nm}$, and varying $L_y$. Fig.~\ref{FIG:Ellipt}a shows the variation of the EDSR Rabi frequency with the aspect ratio $L_x/L_y$, showing that an increase in the aspect ratio results in a larger Rabi frequency. The qubit Zeeman splitting is linear in the applied in-plane magnetic field, similar to the circular case. The relaxation rate varies as $B^3$.
%FROM "In a circular dot...." TO "...relaxation rate varies as $B^3$."; this is my proposed replacement:
%"In circular quantum dots the cubic-symmetry correction $\alpha_{R3}$ is responsible for EDSR, while the spherical term $\alpha_{R2}$ does not contribute. In contrast, in elliptical dots the $\alpha_{R2}$ term is nonzero. From Eqn.~\ref{eq:effectiveSO}, $\alpha_{R3}=(\delta/\overline{\gamma})\,\alpha_{R2}$; using the lattice parameters of Ge we evaluate $\alpha_2{\approx}10\,\alpha_3$. The net SIA Rashba in elliptical dots thus overwhelm the orbital $B_x$ contribution for high ellipticities. Figure~\ref{FIG:Ellipt}a shows the variation of the EDSR Rabi frequency w.r.t. the aspect ratio $L_y/L_x$, where $L_x{=}50$ nm. At high ellipticities of $L_y/L_x\sim 2$ and  $L_y/L_x\sim \frac{1}{2}$, the EDSR Rabi frequency is calculated to show an order of magnitude enhancement. In figures~\ref{FIG:Ellipt}b and \ref{FIG:Ellipt}c we present the results for a high ellipticity QD with dimensions $L_x{=}50\,\text{nm}$, $L_y{=}100\,\text{nm}$, $L_z{=}11\,\text{nm}$. The qubit Zeeman splitting is linear in the applied in-plane magnetic field for such a dot, similar to the circular example. The relaxation rate varies as $B^3$." --> Abhik.
The EDSR Rabi frequency is linear in $B$ (Fig.~\ref{FIG:Ellipt}b), which is reminiscent of out-of-plane $B$-field operation in presence of strong SIA; and the Rabi frequency exhibits a maximum as a function of $F_z$ (Fig.~\ref{FIG:Ellipt}c).

\section{\texorpdfstring{$\bm{g}$}{g}-factor anisotropy of elliptical QD and comparison with experiment}\label{experiment}

The in-plane $g$-factor of an elliptical dot is strongly anisotropic and exhibits an oscillatory behaviour as the magnetic field is rotated in the plane. We compare the predicted variation of the $g$-factor with experimentally measured values from EDSR in a planar germanium hole qubit. The qubit sample is a gate-defined double quantum dot in a Ge/Si$_{0.2}$Ge$_{0.8}$ heterostructure. (See Ref. \citenum{Hendrickx2020} for further sample info). The dots are assumed to be elliptical, with a slight misalignment of their semi-major axes. 
\begin{figure}[htbp!]
    \centering
\includegraphics[width=0.5\textwidth]{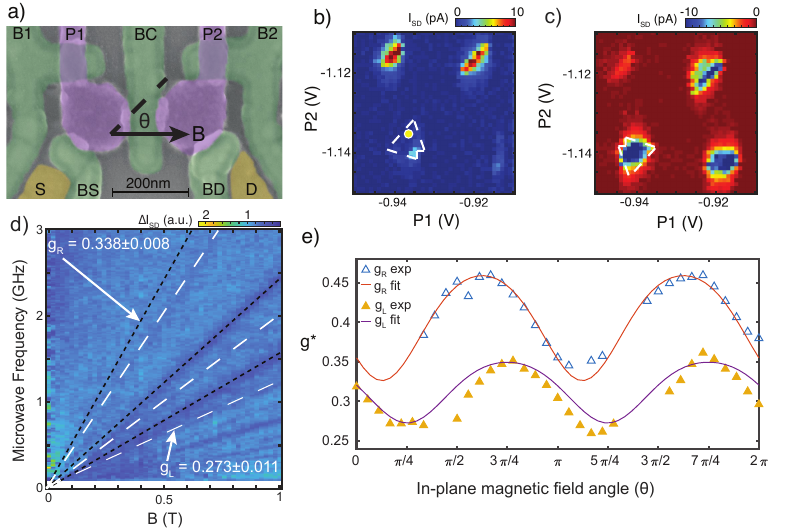}
    \caption{a) Sample scanning electron microscope (SEM) image of a gate-defined Ge double dot. b), c) Bias triangles (dashed) in forward and reverse bias respectively. Pauli spin blockade is visible in the transition marked by a yellow circle. d) g-factors measured via EDSR for the magnetic field direction indicated in part a) (black solid arrow). e) Comparison of the g-factor measured via EDSR with the theoretical prediction. The experimental results for $g$-factor anisotropy in Ref.\citenum{Matthew} are presented as scatter plots of effective $g$-factor ($g^*$) vs. the in-plane magnetic field angular orientation w.r.t. the double dot transport direction ($\theta$). We use triangular markers to denote the experimental data in scatter plots, while the solid fit lines are generated using the theoretical model in Sec.~\ref{sec:Hamil}. The $g$-factor for the left dot is denoted by $g_L$ (empty blue triangular experimental data points, solid red theoretical fit line), while the right dot $g$-factor is denoted by $g_R$ (filled yellow triangular experimental data points, solid purple theoretical fit line).}
    \label{fig:Experiment}
\end{figure}
Fig.~\ref{fig:Experiment}a shows a false colour SEM of the gate design. Plunger gates (purple) are used to define the two dots, while the barrier gates (green) are used to control coupling to the leads and between the dots. Metal ohmic contacts (yellow) act as a reservoir for holes. By negatively biasing the barrier and plunger gates, the sample can be tuned to the few hole regime. The relative angle between the applied magnetic field direction in the plane and the double dot transport direction is denoted by $\theta$. Bias triangles of the double dot measured via transport for positive and negative bias are shown in Fig.~\ref{fig:Experiment}b and Fig.~\ref{fig:Experiment}c. A region of Pauli spin blockade is visible at the base of one charge transition as indicated by a yellow circle. By applying an external magnetic field and a microwave tone to the P2 gate, we are able to drive spin rotations via EDSR when the microwave frequency matches the Larmor frequency ($\hbar f{=}g\mu_B B$). These spin rotations lift the Pauli spin blockade, causing a change in the current through the double dot. Using a lock-in amplifier, we measure the difference in current through the double dot when the microwave is on vs off. Fig.~\ref{fig:Experiment}d shows the change in the leakage current for the double quantum dot with the external magnetic field applied in the direction indicated in Fig.~\ref{fig:Experiment}a. Clear EDSR lines are visible for both dots, and both single and multi photon lines can be seen. From the slope of these resonance lines the g-factor can be calculated for each dot. Using this technique, we measure the g-factor as a function of field angle by rotating a magnetic field of $B{=}0.7\,\text{T}$ in the 2D plane. Fig.~\ref{fig:Experiment}e shows the results of this measurement for both quantum dots, revealing an oscillatory variation in the g-factor as a function of in-plane magnetic field angle. The direction of $\theta$ is shown in Fig.~\ref{fig:Experiment}a.

Using the model developed in Sec.~\ref{sec:Hamil}, we fit to the experimental data. For both dots we use the same size, shape and strain. We are able to account for the difference in g-factor between the dots by considering only a rotation of the dot axes in-plane and a change in the magnitude of the vertical electric field. A full list of fitting parameters is given in Table~\ref{tab:fitparameters}. The maximum value of the g-factor is not aligned with the external magnetic field or sample axes, and is also different for the left and right dots. To account for this, we introduce a phase shift angle ($\theta_{ps}$) which effectively rotates the axes of the quantum dots. Here $\theta_{ps,l}{=}\frac{3\pi}{4}$ and $\theta_{ps,r}{=}\frac{5\pi}{8}$ The magnitude of the g-factor is also different for each dot. This is accounted for by changing the vertical electric field applied to each dot. Here we use 10MV/m for the left dot and 45MV/m for the right dot. The results of the fits for both dots are shown by the solid lines in Fig~\ref{fig:Experiment} e). 

%Previous version - The experimental data further shows that the g-factor of both the left and the right dot is maximum at $\theta>\frac{\pi}{2}$ regime, and that there is a difference of $\theta$ values between where the left and the right dot exhibit g-factor maximum. Using the analytical model developed in Sec.~\ref{sec:Hamil}, we calculate the theoretical g-factor anisotropy and fit to the experimental data. The fitting is shown as solid lines in Fig.~\ref{fig:Experiment}e. We choose similar dot sizes for the left and right QDs: in-plane dot dimensions $40 \text{nm}\times 30 \text{nm}$, and $L_z{=}10 \text{nm}$. Denoting the angular orientation of the semi-major axes of the elliptical dots w.r.t. the double dot transport direction as 'phase-shift' angles $\theta_{ps}$, we input $\theta_{ps,l}{=}\frac{3\pi}{4}$ and $\theta_{ps,r}{=}\frac{5\pi}{8}$ as the phase shift angles for the left and right dots, respectively. WHAT ARE THESE PHASE SHIFT ANGLES AND WHY? I ASSUME THEY DESCRIBE THE ROTATION OF THE ELLIPTICAL QUANTUM DOT AXES WITH RESPECT TO SOMETHING, BUT TO WHAT? PLEASE INSERT A SENTENCE. 
%the previous sentence includes the answers. --> Abhik.

% Phase angle: the elliptical quantum dot with respect to the lab frame -- the growth direction of the crystal?...

\begin{table}[!htbp]
\centering
\begin{tabular}{ l | c  } 
 \hline
 \hline
  Fitting Parameters & values\\
 \hline
 Left QD in-plane dimensions & $40\, \text{nm}{\times}30 \,\text{nm}$ \\
 Right QD in-plane dimensions & $40 \,\text{nm}{\times} 30 \,\text{nm}$ \\
 Left dot perpendicular confinement ($L_{z,l}$) & $10 \,\text{nm}$ \\
 Right dot perpendicular confinement ($L_{z,r}$) & $10 \,\text{nm}$ \\
 Left dot Top-gate voltage ($F_{z,l}$) & $10 \,\text{MV/m}$ \\
 Right dot Top-gate voltage ($F_{z,r}$) & $45\,\text{MV/m}$ \\
 Left dot phase shift $\theta_{ps,l}$ & $\frac{3\pi}{4}$ \\
 Right dot phase shift $\theta_{ps,r}$ & $\frac{5\pi}{8}$ \\
 Left dot uni-axial compressive strain $\varepsilon_{xx,l}$ & $-0.006$ \\
Right dot uni-axial compressive strain $\varepsilon_{xx,r}$ & $-0.006$ \\
Applied magnetic field magnitude $B$ & $0.7\,\text{T}$\\
\hline
\hline
\end{tabular}
\caption{List of the parameters used to fit the analytical model to the experimental $g$-factor anisotropy data in Fig.~\ref{fig:Experiment}e. While the choices of phase shift angles ($\theta_{ps}$) explain the experimental artifact of relative misalignment of the two dots; the top-gate voltages ($F_z$) are tuned to match the $g$-factor oscillation amplitudes of the dots. We stress however that the choice of fitting parameters given in this table for Fig.~\ref{fig:Experiment}e is not unique. A possible set of parameters that yields a similar theoretical fitting of the experimental data in Fig.~\ref{fig:Experiment}e are listed in appendix~\ref{append:anisotropy}.}
\label{tab:fitparameters}
\end{table}

%Our theoretical predictions for both the left dot and the right dot, shown as solid lines in Fig./.\ref{fig:Experiment}e), use fitting parameter $L_x$ = 30 nm, $L_y$ = 40 nm, $L_z$ = 10 nm, $\varepsilon_{xx}$ = -0.006 for both dots. To account for the difference of the $g$-factor oscillation amplitudes, we change the vertical electric field, fitting F = 10 MV/m for the right dot, F = 45 MV/m for the left dot. In addition, there is a difference in the angle of the maximum $g$-factor for each dot; in the experiments, this is most likely to be from the gate lithography that used to define the quantum dot. In theory fittings, we rotate our magnetic field effectively for two quantum dots, by introducing a phase shift angle. The phase shift angle for the right dot is $3\pi/4$, and for the left dot it is $5\pi/8$. This demonstrates that our model can capture the $g$-factor anisotropy of hole quantum dots in Ge by changing the vertical electric field and rotating the dot. 

 The theoretical fit in Fig.~\ref{fig:Experiment}e shows good agreement between the phase shift angle $\theta_{ps}$ parameter choices and B-field angular orientations where the experimental $g$-factors are maximum. In other words, the largest value of the $g$-factor occurs when the magnetic field is parallel to the semi-major axis of the elliptical hole QD. This behaviour is consistent with the effective in-plane $g$-factor being primarily the result of coupling to higher excited states brought about by the orbital magnetic field terms. We note that inhomogeneous strain in the sample, or the Ge/SiGe heterostructure interface induced roughness and disorder, or a misalignment of the sample with respect to the in-plane B-field (since $g_\perp\gg g_\parallel$) could potentially lead to significant modulation of the $g$-tensor. We can rule out the latter, since there is a different phase shift for the left and right dots in Fig.~\ref{fig:Experiment}e. The effects of strain and inhomogeneities on $g$-factor anisotropy will be considered in a future publication.

% When the in-plane magnetic field aligns with the semi-major axis, our numerical simulations show that the RSOC corrections due to lateral confinement are maximized as well. The qubit Zeeman splitting, which depends on the overlap between the ground state wavefunction and higher excited states via all the orbital terms, can be traced back to the non-commutative wavevector operators. The increase in dot size leads to an increase in wavefunction overlap, resulting in larger g-factors if there is a semi-major axis in the quantum dot.

\section{Conclusions and Outlook}\label{sec:outlook}

We have presented a generalised, semi-analytical model that fully describes the electrical operation of a planar germanium hole qubit in presence of an in-plane magnetic field. A comprehensive theory for spin manipulation via electron dipole spin resonance (EDSR) is given: surface inversion asymmetry (SIA) mediated fast EDSR is a result of the $\mathbf{k}\cdot\mathbf{p}$ coupling of the heavy hole ground state to higher energy light hole bands. The EDSR rate is linear in $B$ with important nonlinear corrections due to orbital mixing. Qubit relaxation is induced by acoustic phonons and the relaxation rate $1/T_1$ has terms with $B^3,\,B^4,\,B^5$ dependence, again reflecting the importance of the orbital mixing. In-plane operation demonstrates an excellent trade-off between relaxation and EDSR. The in-plane $g$-factor is strongly anisotropic and oscillates as the magnetic field is rotated in the plane. Random telegraph noise from charges in the plane of the quantum well results in decoherence, with an optimal top gate potential where it is insensitive to $\Delta F_z$; although the in-plane magnetic field exposes the qubit to $x-y$ electric field of the fluctuator. Hence, in contrast to the case of out-of-plane magnetic fields, coherence sweet spots cannot be identified in an in-plane $B$ for a qubit exposed to electric field fluctuations in all spatial directions. For an elliptical QD of aspect ratio $L_y/L_x{=}2$, EDSR is shown to be faster by an order of magnitude compared to a circular dot of $L_x{=}50\,\text{nm}$ radius; and the non-linear correction to EDSR is suppressed as rotational asymmetry induces more SIA Rashba. 

 %The actual electrostatic landscape of the Ge hole quantum dot is highly non-trivial. The $z$-component of the electric field and the potential at the interface are almost certainly non-uniform across the dot. This leads to a finite probability that the light hole states with higher energies may leak into the SiGe barrier, or that Fowler-Nordheim tunnelling of the heavy hole states will take place from the quantum well into the barrier; the consequences of which are beyond the hard-wall potential model considered in this work. The potential profile can change due to charges being trapped at the semiconductor/ALD oxide interface. Moreover, non-uniform strain can lead to a significant modification of the spin-orbit interaction. Further progress must rely on sophisticated numerical simulation techniques, including detailed 3D COMSOL modelling.
 
\textit{Acknowledgments}. This project is supported by the Australian Research Council Centre of Excellence in Future Low-Energy Electronics Technologies (project number CE170100039) and Discovery Project DP200100147. We acknowledge stimulating discussions with S. Das Sarma, M. Russ, X. Hu, and S. Liles. 

%\bibliography{ReferenceAll}
%\bibliography{ReferenceAll}

\clearpage
\renewcommand{\theequation}{\thesection\arabic{equation}}
\renewcommand{\thefigure}{\thesection\arabic{figure}}
\renewcommand{\bibnumfmt}[1]{[S#1]}

\appendix

\begin{widetext}
\section{Gauge Invariance}
The hole motion in the topmost valence band is described by the $4{\times}4$ Luttinger-Kohn (LK) Hamiltonian, which in the general operator form is given by:
  \begin{eqnarray}
      H_{LK}&=&\frac{\hbar^2}{2m_0}\left[\left(\gamma_1+\frac{5\gamma_2}{2}\right)k^2-2 \gamma_2(k_x^2J_x^2+k_y^2J_y^2+k_z^2J_z^2)-4\gamma_3\left(\{k_x,k_y\}\{J_x,J_y\}+c.p.\right)\right]
  \end{eqnarray}
  Expanding the anti-commutators, LK Hamiltonian has the form:
  \begin{eqnarray}
      H_{LK}&=&\frac{\hbar^2}{2m_0}\left[\left(\gamma_1+\frac{5\gamma_2}{2}\right)(k_x^2+k_y^2+k_z^2)-2 \gamma_2(k_x^2J_x^2+k_y^2J_y^2+k_z^2J_z^2)-4\gamma_3\left(\left(\frac{k_xk_y+k_yk_x}{2}\right)\left(\frac{J_xJ_y+J_yJ_x}{2}\right)\right.\right.\nonumber\\
      &&+\left(\frac{k_yk_z+k_zk_y}{2}\right)\left(\frac{J_yJ_z+J_zJ_y}{2}\right)\left.\left.+\left(\frac{k_xk_z+k_zk_x}{2}\right)\left(\frac{J_xJ_z+J_zJ_x}{2}\right)\right)\right]
  \end{eqnarray}
  with $m_0$ notifying bare electron mass, $\gamma_1{=}13.38$, $\gamma_2{=}4.24$ and $\gamma_3{=}5.69$ are Luttinger parameters. 
We have tried two different gauges: $\frac{1}{2}\mathbf{B}\times\mathbf{r}$ and gauge choice from Loss et al.\cite{kloeffel2018direct} 

\subsection{Gauge 1:} 
In presence of magnetic field, the momentum correction would be: $\mathbf{k}\rightarrow\left(\mathbf{k}+\frac{e\mathbf{A}}{\hbar}\right)$. We use the gauge (symmetric in z, Landau in x-y) $\mathbf{A}=-\frac{1}{2}B_z y\mathbf{\hat{e}_x}+\frac{1}{2}B_z x\mathbf{\hat{e}_y}+(B_x y-B_y x)\mathbf{\hat{e}_z}$.

\textbf{Momentum Correction:}\begin{enumerate}
    \item The components of corrected momentum:
   \begin{eqnarray}
       k_x &\rightarrow& \left(k_x-\frac{eB_z}{2\hbar}y\right)\, ,\, k_y\rightarrow \left(k_y+\frac{eB_z}{2\hbar}x\right)\, ,\,
         k_z \rightarrow\left(k_z+\frac{eB_x}{\hbar}y-\frac{eB_y}{\hbar}x\right)\nonumber\\
         \nonumber
   \end{eqnarray}
    where, $\mathbf{k}=-\iota\mathbf{\partial}$. For $B_y=0,\,B_z=0$; $\text{we write: } \,k_x \rightarrow k_x\, ,\, k_y\rightarrow k_y\, ,\,k_z \rightarrow \left(k_z+\frac{eB_x}{\hbar}y\right)$
   \item We evaluate $k_i^2$ terms below:
   \begin{eqnarray}
       k_x^2&\Rightarrow& k_x^2\, ,\,
       k_y^2\Rightarrow k_y^2\, ,\,k_z^2\Rightarrow\left(k_z^2+\frac{2eB_xyk_z}{\hbar}+\frac{e^2B_x^2y^2}{\hbar^2}\right)
   \end{eqnarray}
   and the cross-terms $k_ik_j$ are:
   \begin{eqnarray}
    k_xk_y\Rightarrow k_xk_y;\,k_yk_z\Rightarrow\left(k_yk_z+\frac{eB_x}{\hbar}\{y,k_y\}\right);\,k_xk_z\Rightarrow\left(k_xk_z+\frac{eB_xyk_x}{\hbar}\right)
\end{eqnarray}

\item Using these corrections, the LK Hamiltonian is:
\begin{eqnarray}
      &&H_{LK}=\bigg[\,\overline{\gamma_1+\frac{5\gamma_2}{2}}\left(k_x^2+k_y^2+k_z^2+\frac{2eB_xyk_z}{\hbar}+\frac{e^2B_x^2y^2}{\hbar^2}\right)-2\gamma_2\left(k_x^2J_x^2+k_y^2J_y^2+\left(k_z^2+\frac{2eB_xyk_z}{\hbar}+\frac{e^2B_x^2y^2}{\hbar^2}\right)J_z^2\right)\nonumber\\
      &&-4\gamma_3\left(k_xk_y \{J_x,J_y\}+\left(k_yk_z+\frac{eB_x}{\hbar}\{y,k_y\}\right)\{J_y,J_z\}+\left(k_xk_z+\frac{e B_x y k_x}{\hbar}\right)\{J_z,J_x\}\right)\bigg]\frac{\hbar^2}{2m_0}
  \end{eqnarray}
where,
\begin{eqnarray}
   H_{LK}^{11/22} &=& \frac{\hbar^2}{2m_0}\bigg[(\gamma_1-2\gamma_2)\left(k_z^2+\frac{2eB_xyk_z}{\hbar}+\frac{e^2B_x^2y^2}{\hbar^2}\right)+(\gamma_1+\gamma_2)(k_x^2+k_y^2)\bigg]\nonumber\\
   H_{LK}^{33/44} &=& \frac{\hbar^2}{2m_0}\bigg[(\gamma_1+2\gamma_2)\left(k_z^2+\frac{2eB_xyk_z}{\hbar}+\frac{e^2B_x^2y^2}{\hbar^2}\right)+(\gamma_1-\gamma_2)(k_x^2+k_y^2)\bigg]\nonumber\\
   H_{LK}^{13}=L&=& -\sqrt{3}\frac{\hbar^2\gamma_3}{m_0}\bigg[\left(k_xk_z+\frac{eB_xyk_x}{\hbar}\right)-i\left(k_yk_z+\frac{eB_x}{\hbar}\{y,k_y\}\right)\bigg]\nonumber\\
   H_{LK}^{14}=M&=& \frac{\sqrt{3}\hbar^2}{2m_0}\left[-\gamma_2 \left(k_x^2-k_y^2\right)+2i\gamma_3 k_xk_y\right]\nonumber\\
\end{eqnarray}
\end{enumerate}
\setcounter{figure}{0}
\begin{figure}
%\subfloat[]{\begin{minipage}[c]{0.5\linewidth}
\includegraphics[width=3.5 in, height=2 in]{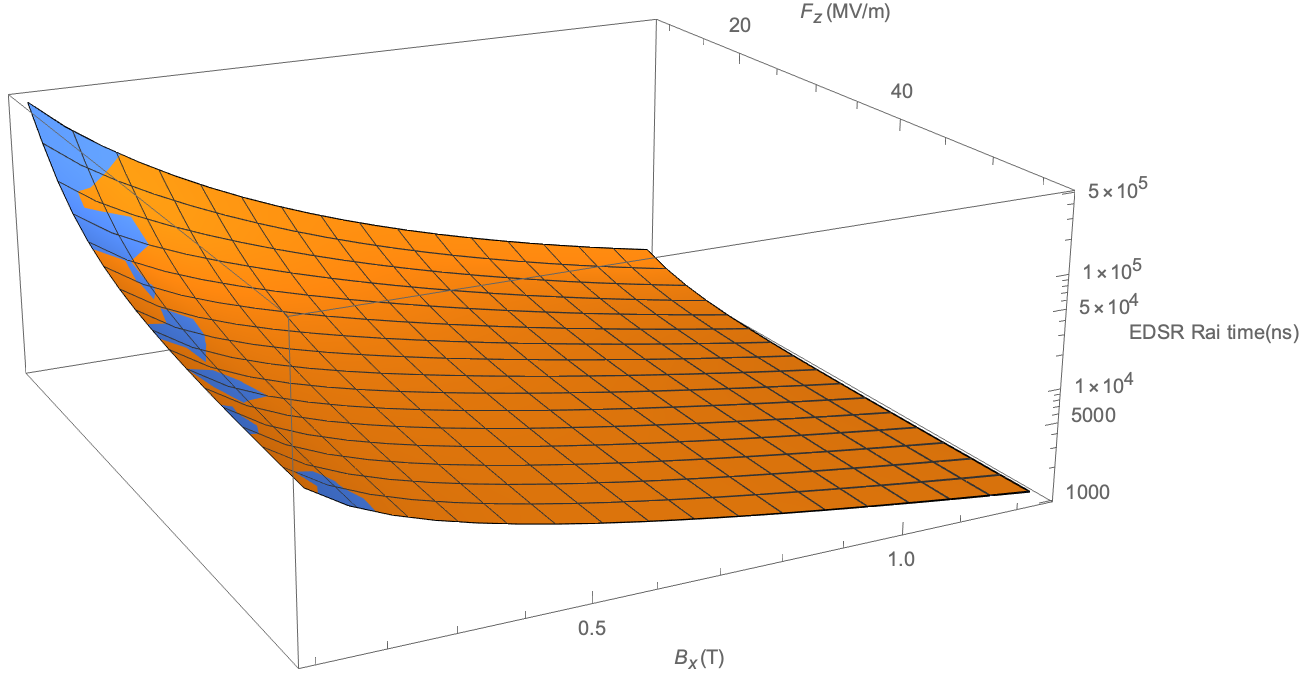}
%\end{minipage}}
\caption{EDSR time $T_\pi$ (ns) vs. top gate field $F_z$ and in-plane magnetic field $B_x$.}
	\label{fig:EDSRrate3D}
\end{figure}
\subsection{Gauge 2:} 
In this section, we use the symmetric gauge ($\frac{1}{2}\mathbf{B}\times\mathbf{r}$): $\mathbf{A}=\frac{1}{2}(B_yz-B_zy)\mathbf{\hat{e}_x}+\frac{1}{2}(B_zx-B_xz)\mathbf{\hat{e}_y}+\frac{1}{2}(B_x y-B_y x)\mathbf{\hat{e}_z}$.

\textbf{Momentum Correction:}\begin{enumerate}
    \item The components of corrected momentum with $B_y=0\,,B_z=0$:
   \begin{eqnarray}
       k_x \rightarrow k_x ,\, k_y\rightarrow \left(k_y-\frac{eB_x}{2\hbar}z\right)\, ,\,
         k_z \rightarrow \left(k_z+\frac{eB_x}{2\hbar}y\right)\,\text{where,}\,\mathbf{k}=-\iota\partial.
   \end{eqnarray}
   \item We evaluate $k_i^2$ terms below:
   \begin{eqnarray}
       k_x^2&\Rightarrow& k_x^2\, ,\,
       k_y^2\Rightarrow \left(k_y^2-\frac{eB_xzk_y}{\hbar}+\frac{e^2B_x^2z^2}{4\hbar^2}\right)\, ,\,k_z^2\Rightarrow\left(k_z^2+\frac{eB_xyk_z}{\hbar}+\frac{e^2B_x^2y^2}{4\hbar^2}\right)
   \end{eqnarray}
   and the cross-terms $k_ik_j$ are:
   \begin{eqnarray}
   k_xk_y&\Rightarrow&\!\left(\!k_xk_y-\frac{eB_xzk_x}{2\hbar}\!\right);\!k_yk_z\Rightarrow\!\left(\!k_yk_z+\frac{eB_x}{2\hbar}(\{y,k_y\}-\{z,k_z\})-\frac{e^2B_x^2}{4\hbar^2}yz\right)\!;\!k_xk_z\Rightarrow\left(\!k_xk_z+\frac{eB_xyk_x}{2\hbar}\right)
\end{eqnarray}

 \item Using these corrections, the LK Hamiltonian terms become:
 \begin{eqnarray}
      &&H_{LK}=\frac{\hbar^2}{2m_0}\bigg[\left(\gamma_1+\frac{5\gamma_2}{2}\right)\left(k_x^2+k_y^2-\frac{eB_xzk_y}{\hbar}+\frac{e^2B_x^2z^2}{4\hbar^2}+k_z^2+\frac{eB_xyk_z}{\hbar}+\frac{e^2B_x^2y^2}{4\hbar^2}\right)\nonumber\\
&&-2\gamma_2\left(k_x^2J_x^2+\left(k_y^2-\frac{eB_xzk_y}{\hbar}+\frac{e^2B_x^2z^2}{4\hbar^2}\right)J_y^2+\left(k_z^2+\frac{eB_xyk_z}{\hbar}+\frac{e^2B_x^2y^2}{4\hbar^2}\right)J_z^2\right)-4\gamma_3\left(\left(k_xk_y-\frac{eB_xzk_x}{2\hbar}\right) \{J_x,J_y\}\right.\nonumber\\
      &&\left.+\left(k_yk_z+\frac{eB_x}{2\hbar}\{y,k_y\}-\frac{eB_x}{2\hbar}\{z,k_z\}-\frac{e^2B_x^2}{4\hbar^2}yz\right)\{J_y,J_z\}+\left(k_xk_z+\frac{eB_xyk_x}{2\hbar}\right)\{J_z,J_x\}\right)\bigg]
  \end{eqnarray}
  where,
\begin{eqnarray}
  && H_{LK}^{11/22} = \frac{\hbar^2}{2m_0}\left[(\gamma_1-2\gamma_2)\left(k_z^2+\frac{2eB_xyk_z}{\hbar}+\frac{e^2B_x^2y^2}{\hbar^2}\right)+(\gamma_1+\gamma_2)\left(k_x^2+k_y^2-\frac{eB_xzk_y}{\hbar}+\frac{e^2B_x^2z^2}{4\hbar^2}\right)\right]
   \end{eqnarray}
\begin{eqnarray}
   &&H_{LK}^{33/44} = \frac{\hbar^2}{2m_0}\left[(\gamma_1+2\gamma_2)\left(k_z^2+\frac{2eB_xyk_z}{\hbar}+\frac{e^2B_x^2y^2}{\hbar^2}\right)+(\gamma_1-\gamma_2)\left(k_x^2+k_y^2-\frac{eB_xzk_y}{\hbar}+\frac{e^2B_x^2z^2}{4\hbar^2}\right)\right]
   \end{eqnarray}
\begin{eqnarray}
   &&H_{LK}^{13}=L= -\sqrt{3}\frac{\hbar^2\gamma_3}{m_0}\bigg[\left(k_xk_z+\frac{eB_xyk_x}{2\hbar}\right)-i\left(k_yk_z+\frac{eB_x}{2\hbar}\{y,k_y\}-\frac{eB_x}{2\hbar}\{z,k_z\}-\frac{e^2B_x^2}{4\hbar^2}yz\right)\bigg]\nonumber\\
  && H_{LK}^{14}=M= \frac{\sqrt{3}\hbar^2}{2m_0}\bigg[-\gamma_2 \left(k_x^2-k_y^2+\frac{eB_xzk_y}{\hbar}-\frac{e^2B_x^2z^2}{4\hbar^2}\right)+2i\gamma_3 \left(k_xk_y-\frac{eB_xzk_x}{2\hbar}\right)\bigg]
\end{eqnarray}
\end{enumerate}

\section{Schreiffer-Wolff Transformation}\label{appen:EDSR}

In the effective $2\times2$ system Hamiltonian given by Eqn.~\ref{eq:effective2DEDSR}, $H_0+V(x,y)$ produces the quantized in-plane energies. We consider a basis with following in-plane states:
\begin{equation}
    \left\{\frac{e^{-\left(\!\frac{x^2}{2L_x^2}{+}\frac{y^2}{2L_y^2}\right)}}{\sqrt{\pi L_x L_y}},\,\frac{\sqrt{2}y\,e^{-\left(\!\frac{x^2}{2L_x^2}{+}\frac{y^2}{2L_y^2}\right)}}{\sqrt{\pi L_y^3 L_x}},\,\frac{\sqrt{2}x\,e^{-\left(\!\frac{x^2}{2L_x^2}{+}\frac{y^2}{2L_y^2}\right)}}{\sqrt{\pi L_x^3 L_y}}\right\}\otimes\left\{\uparrow,\,\downarrow\right\}
\end{equation}
where the spatial states are the first three harmonic oscillator product states: $(n,m){=}\{(0,0),(0,1),(1,0)\}$, we denote the orbital energies of these states as $E_0,\,E_1,\,E_1$. The spinors are the effective spin-up and spin-down states of the 2D hole qubit, and gives $\pm g_\parallel\mu_BB_x$ Zeeman energies for the up and down spin states. The spin-orbit interactions can be listed as:
\begin{equation}
    H_{SO}{=}i\alpha_{R1}(k_-\sigma_-{-}k_+\sigma_+)\!+i\alpha_{R2}\left(k_+^3\sigma_--k_-^3\sigma_+\right)\!+i\alpha_{R3}\left(\{k_+,k_-^2\}\sigma_-\!-\!\{k_+^2,k_-\}\sigma_+\right)\!+i\alpha_{R4}\left(\{k_z^2,k_-\}\sigma_-{-}\{k_z^2,k_+\}\sigma_+\right)
\end{equation}
The $k$-linear Rashba terms come from the coupling of the bonding VB $p$-orbitals to the antibonding CB $p$-orbitals, which is very small. In the Luttinger formalism thus the $\alpha_{R1},\,\alpha_{R4}$ terms vanish. The nonzero contributions are:
\begin{equation}
    H_{SO}{=}i\alpha_{R2}\left(k_+^3\sigma_--k_-^3\sigma_+\right)\!+i\alpha_{R3}\left(\{k_+,k_-^2\}\sigma_-\!-\!\{k_+^2,k_-\}\sigma_+\right)
\end{equation}
the spin-orbit matrix elements are calculated below:
\begin{eqnarray}\label{eqn:appencomplexRabi}
    \left\langle(0,0)\uparrow\right|H_{SO}\left|(0,1)\downarrow\right\rangle&{=}&-i\alpha_{R2}\left\langle(0,0)\uparrow\right|k_-^3\sigma_+\left|(0,1)\downarrow\right\rangle+i\alpha_{R3}\left\langle(0,0)\uparrow\right|k_+k_-k_+\sigma_+\left|(0,1)\downarrow\right\rangle\nonumber\\
    &{=}&-i\alpha_{R2}\left\langle(0,0)\right|k_-^3\left|(0,1)\right\rangle+i\alpha_{R3}\left\langle(0,0)\right|k_+k_-k_+\left|(0,1)\right\rangle\nonumber\\
    &{=}&\frac{-3i\alpha_{R2}}{2\sqrt{2}L_y^3}\left(1-\frac{L_y^2}{L_x^2}\right)+\frac{i\alpha_{R3}}{2\sqrt{2}L_y^3}\left(1+3\frac{L_y^2}{L_x^2}\right)=R_c
\end{eqnarray}
\begin{eqnarray}\label{eqn:appenrealRabi}
    \left\langle(0,0)\uparrow\right|H_{SO}\left|(1,0)\downarrow\right\rangle&{=}&-i\alpha_{R2}\left\langle(0,0)\uparrow\right|k_-^3\sigma_+\left|(1,0)\downarrow\right\rangle+i\alpha_{R3}\left\langle(0,0)\uparrow\right|k_+k_-k_+\sigma_+\left|(1,0)\downarrow\right\rangle\nonumber\\
    &{=}&-i\alpha_{R2}\left\langle(0,0)\right|k_-^3\left|(0,1)\right\rangle+i\alpha_{R3}\left\langle(0,0)\right|k_+k_-k_+\left|(0,1)\right\rangle\nonumber\\
    &{=}&\frac{-3\alpha_{R2}}{2\sqrt{2}L_xL_y^2}\left(1-\frac{L_y^2}{L_x^2}\right)+\frac{\alpha_{R3}}{2\sqrt{2}L_xL_y^2}\left(1+3\frac{L_y^2}{L_x^2}\right)=R_r
\end{eqnarray}
The ac electric field $eE_x(t)x$ is spin-conserving and connects the states $(0,0)$ and $(1,0)$. Putting all of it together, the $6\times6$ Hamiltonian would be:
\begin{equation}
    H'=\begin{pmatrix}
        E_0-\frac{1}{2}g_\parallel\mu_BB_x & 0 & 0 & R_c & eE_xa & R_r
\\
0 &  E_0+\frac{1}{2}g_\parallel\mu_BB_x & R_c^* & 0 & -R_r & eE_xa\\
0 & R_c & E_1-\frac{1}{2}g_\parallel\mu_BB_x & 0 & 0 & 0\\
R_c^* & 0 & 0 & E_1+\frac{1}{2}g_\parallel\mu_BB_x & 0 & 0\\
eE_xa & -R_r & 0 & 0 & E_1-\frac{1}{2}g_\parallel\mu_BB_x & 0\\
R_r & eE_xa & 0 & 0 & 0 & E_1+\frac{1}{2}g_\parallel\mu_BB_x\end{pmatrix}
\end{equation}
 $2^\text{nd}$ order perturbation theory gives the EDSR matrix element as:
 \begin{eqnarray}
     \tilde{H}_{12}&=&\sum_{j=3}^6\frac{1}{2}H'_{1j}H'_{j2}\left(\frac{1}{\epsilon_1-\epsilon_j}+\frac{1}{\epsilon_2-\epsilon_j}\right)=\frac{1}{2}H'_{15}H'_{52}\left(\frac{1}{\epsilon_1-\epsilon_5}+\frac{1}{\epsilon_2-\epsilon_5}\right)+\frac{1}{2}H'_{16}H'_{62}\left(\frac{1}{\epsilon_1-\epsilon_6}+\frac{1}{\epsilon_2-\epsilon_6}\right)\nonumber\\
     &=&-\frac{1}{2}eE_xa\,R_r\left(\frac{1}{-\Delta_{01}}+\frac{1}{-\Delta_{01}+Z}\right)+\frac{1}{2}eE_xa\,R_r\left(\frac{1}{-\Delta_{01}-Z}+\frac{1}{-\Delta_{01}}\right)\nonumber\\
     &=&eE_xa\,R_r\left(\frac{1}{\Delta_{01}\left(1-\frac{Z}{\Delta_{01}}\right)}-\frac{1}{\Delta_{01}\left(1+\frac{Z}{\Delta_{01}}\right)}\right)=(eE_xa\,R_r Z)/\Delta_{01}^2
 \end{eqnarray}
  Where we have used $Z=g_\parallel\mu_BB_x$, $\Delta_{01}=E_1-E_0$. This gives $ \tilde{H}_{12}{\propto}B_x$, explaining the linear magnetic field dependence of EDSR. 

  From Eqn.~\ref{eqn:appencomplexRabi}, \ref{eqn:appenrealRabi}, the $\alpha_2$ term vanishes when $L_x=L_y$ i.e. for a symmetric dot, while it is strongly nonzero for an elliptical dot; making the linear $B_x$-dependence stronger.
%The $k$-linear Rashba terms come from the coupling of the bonding VB $p$-orbitals to the antibonding CB $p$-orbitals, which is very small. In the Luttinger formalism thus the $r_1,\,r_4$ terms vanish. One should note that the explanation here is for a planar hole QD qubit, which is heavy-hole (HH) in nature. For low-symmetry growth directions, or in the prsence of strain, light hole (LH) states exhibit strong linear Rashba coefficients, and this feature has been explored in novel hole qubits with LH ground states.\cite{hu2012hole,ares2013sige,terrazos2021theory}
\section{Bulk Phonon mediated Relaxation Rate Analytics}\label{appen:relaxation}
The total strain in the $4\times4$ Luttinger-Kohn-Bir-Pikus formalism is given as:
\begin{equation}
    H^{LKBP}_{strain}(\bm r)=\begin{pmatrix}
    P_\varepsilon(\bm r)+Q_\varepsilon(\bm r) & 0 & L_\varepsilon(\bm r) & M_\varepsilon(\bm r)\\
    0 & P_\varepsilon(\bm r)+Q_\varepsilon(\bm r) & M_\varepsilon(\bm r)^* & -L_\varepsilon(\bm r)^* \\
    L_\varepsilon(\bm r)^* & M_\varepsilon(\bm r) & P_\varepsilon(\bm r)-Q_\varepsilon(\bm r) & 0 \\
    M_\varepsilon(\bm r)^* & -L_\varepsilon(\bm r) & 0 & P_\varepsilon(\bm r)-Q_\varepsilon(\bm r) \\
    \end{pmatrix}
\end{equation}
where, $P_\varepsilon(\bm r){=}-a(\varepsilon_{xx}(\bm r)+\varepsilon_{yy}(\bm r)+\varepsilon_{zz}(\bm r))$, $Q_\varepsilon(\bm r){=}-\frac{b}{2}(\varepsilon_{xx}(\bm r)+\varepsilon_{yy}(\bm r)-2\varepsilon_{zz}(\bm r))$, $L_\varepsilon(\bm r){=}d(\varepsilon_{xz}(\bm r)-i \varepsilon_{yz}(\bm r))$, $M_\varepsilon(\bm r){=}\left(\frac{\sqrt{3}}{2}b(\varepsilon_{xx}(\bm r)-\varepsilon_{yy}(\bm r))-id\varepsilon_{xy}(\bm r)\right)$. The non-zero static component of the strain tensor is given by $\varepsilon_{xx}=\varepsilon_{yy}=-0.006$, and $\varepsilon_{zz}=-\frac{C_{12}}{C_{11}}\varepsilon_{xx}$ with $C_{12}{=}44$ GPa, $C_{11}{=}126$ GPa. Considering the lattice deformation potential $\mathcal{D}(\bm r)$, the 'local' strain is given by:
\begin{equation}
    \varepsilon_{i,j}^\alpha(\bm r)=\frac{1}{2}\left(\frac{\partial u_i(\bm r)}{\partial r_j}+\frac{\partial u_j(\bm r)}{\partial r_i}\right),\ \    i,j\in\{x,y,z\}
\end{equation}
Here $\bm u$ vector designates the deformation field at the position $\bm r$. For a phonon traveling with wave vector $\mathbf{q}$ in the polarized state $\alpha$, the strain tensor is given by,
\begin{eqnarray}
    \varepsilon_{i,j}^\alpha(\bm r)=\frac{i}{2}\sqrt{\frac{\hbar}{2V_c\rho \omega_{\mathbf{q},\alpha}}}q(\hat{c}_i^\alpha\hat{q}_j+\hat{c}_j^\alpha\hat{q}_i)(e^{-i\mathbf{q}\cdot\mathbf{r}}a_{\mathbf{q},\alpha}+e^{i\mathbf{q}\cdot\mathbf{r}}a^\dagger_{\mathbf{q},\alpha})
\end{eqnarray}
where $\hat{c}$ is the polarisation unit vector. We consider three polarisations with con-ordinate systems understood as $l\!:\!(r,\theta,\phi);\,t\!:\!\left(r,\theta+\frac{\pi}{2},\phi\right);\,w\!:\!\left(r,\frac{\pi}{2},\phi+\frac{\pi}{2}\right)$. Using $\omega_{\mathbf{q},\alpha}{=}\,v_\alpha q$,
\begin{equation}
    \varepsilon_{i,j}^\alpha(\bm r)=i\sqrt{\frac{\hbar}{2V_c\rho v_{\alpha}}}\sqrt{q}\mathcal{A}_{\varepsilon,ij}^\alpha e^{i\mathbf{q}\cdot\mathbf{r}}(a_{-\mathbf{q},\alpha}+a^\dagger_{\mathbf{q},\alpha})
\end{equation}
where $v_\alpha$ are the acoustic phonon velocities, and we assumed $\frac{1}{2}(\hat{c}_i^\alpha\hat{q}_j+\hat{c}_j^\alpha\hat{q}_i)=\mathcal{A}_{\varepsilon,ij}^\alpha$. The matrix elements of $\mathcal{A}_{\varepsilon,ij}^\alpha$ are sketched out below:
\begin{equation}
    \mathcal{A}_\varepsilon^\alpha=\frac{1}{2}\begin{pmatrix}
        2\hat{c}_x^\alpha\hat{q}_x & \hat{c}_x^\alpha\hat{q}_y+\hat{c}_y^\alpha\hat{q}_x & \hat{c}_x^\alpha\hat{q}_z+\hat{c}_z^\alpha\hat{q}_x \\
        \hat{c}_y^\alpha\hat{q}_x+\hat{c}_x^\alpha\hat{q}_y & 2\hat{c}_y^\alpha\hat{q}_y & \hat{c}_y^\alpha\hat{q}_z+\hat{c}_z^\alpha\hat{q}_y \\
        \hat{c}_z^\alpha\hat{q}_x+\hat{c}_x^\alpha\hat{q}_z & \hat{c}_z^\alpha\hat{q}_y+\hat{c}_y^\alpha\hat{q}_z & 2\hat{c}_z^\alpha\hat{q}_z
    \end{pmatrix}
\end{equation}
The phonon wave vector has the components $\overrightarrow{q}\rightarrow\{q \sin\theta\cos\phi,q \sin\theta\, \sin\phi,q \cos\theta\}$; so the unit vectors for $\bm q$ are given by: $\bm \hat{q}\rightarrow\{\sin\theta\cos\phi,\,\sin\theta\, \sin\phi,\,\cos\theta\}$. The polarization wave vectors are as follows:
\begin{itemize}
    \item $\hat{c}^l\rightarrow q^{-1}\{q_x,q_y,q_z\}=\{\sin\theta\cos\phi,\,\sin\theta\cos\phi,\,\cos\theta\}$
    \item $\hat{c}^t\rightarrow q^{-1}(q_x^2+q_y^2)^{-\frac{1}{2}}\{q_xq_z,q_yq_z,-(q_x^2+q_y^2)\}=\{\cos\theta\cos\phi,\,\cos\theta\sin\phi,-\sin\theta\}$
    \item $\hat{c}^w\rightarrow (q_x^2+q_y^2)^{-\frac{1}{2}}\{q_y,-q_x,0\}=\{-\sin\phi,\,\cos\phi,\,0\}$
\end{itemize}
We can write the matrix elements of $\mathcal{A}_{\varepsilon,ij}^\alpha$ for the three polarisations using the decompositions above:
\begin{eqnarray}
    \mathcal{A}_\varepsilon^l&=&\frac{1}{2}\begin{pmatrix}
        2 \frac{q_x^2}{q^2} & 2 \frac{q_xq_y}{q^2} & 2 \frac{q_xq_z}{q^2} \\
        2 \frac{q_yq_x}{q^2} & 2 \frac{q_y^2}{q^2} & 2 \frac{2 q_yq_z}{q^2} \\
        2 \frac{q_zq_x}{q^2} & 2 \frac{q_zq_y}{q^2} & 2 \frac{q_z^2}{q^2}
    \end{pmatrix}=\frac{1}{q^2}\begin{pmatrix}
        q_x^2 & q_xq_y & q_xq_z \\
        q_yq_x & q_y^2 & q_yq_z \\
        q_zq_x & q_zq_y & q_z^2 \\
    \end{pmatrix}\nonumber
    \end{eqnarray}
    \begin{eqnarray}
    \mathcal{A}_\varepsilon^t&=&\frac{1}{2}\begin{pmatrix}
        2 \frac{q_z}{q}\frac{q_x}{\sqrt{q_x^2+q_y^2}}\frac{q_x}{q} & 2 \frac{q_z}{q}\frac{q_x}{\sqrt{q_x^2+q_y^2}}\frac{q_y}{q} & \frac{q_xq_z}{q\sqrt{q_x^2+q_y^2}}\frac{q_z}{q}-\frac{\sqrt{q_x^2+q_y^2}}{q}\frac{q_x}{q} \\
        2 \frac{q_z}{q}\frac{q_x}{\sqrt{q_x^2+q_y^2}}\frac{q_y}{q} & 2 \frac{q_z}{q}\frac{q_y}{\sqrt{q_x^2+q_y^2}}\frac{q_y}{q} & \frac{q_yq_z}{q\sqrt{q_x^2+q_y^2}}\frac{q_z}{q}-\frac{\sqrt{q_x^2+q_y^2}}{q}\frac{q_y}{q}\\
        \frac{q_xq_z}{q\sqrt{q_x^2+q_y^2}}\frac{q_z}{q}-\frac{\sqrt{q_x^2+q_y^2}}{q}\frac{q_x}{q} & \frac{q_z^2q_y-q_x^2q_y-q_y^3}{q^2\sqrt{q_x^2+q_y^2}} & -2\frac{q_z}{q}\frac{\sqrt{q_x^2+q_y^2}}{q}
    \end{pmatrix}\nonumber
\end{eqnarray}
\begin{eqnarray}
    \mathcal{A}_\varepsilon^w&=&\frac{1}{2}\begin{pmatrix}
        -2\frac{q_y}{\sqrt{q_x^2+q_y^2}}\frac{q_x}{q} & -\frac{q_y}{\sqrt{q_x^2+q_y^2}}\frac{q_y}{q}+\frac{q_x}{\sqrt{q_x^2+q_y^2}}\frac{q_x}{q} & -\frac{q_y}{\sqrt{q_x^2+q_y^2}}\frac{q_z}{q}\\
        \frac{q_x^2-q_y^2}{q\sqrt{q_x^2+q_y^2}} & 2\frac{q_xq_y}{q\sqrt{q_x^2+q_y^2}} & \frac{q_xq_z}{q\sqrt{q_x^2+q_y^2}}\\
        -\frac{q_yq_z}{q\sqrt{q_x^2+q_y^2}} & \frac{q_xq_z}{q\sqrt{q_x^2+q_y^2}} & 0
    \end{pmatrix}
\end{eqnarray}
We wish to evaluate the angular integrals, so we write the matrices in terms of $\theta$ and $\phi$:
\begin{equation}
    \mathcal{A}_\varepsilon^l=\begin{pmatrix}
        \sin^2\theta\cos^2\phi & \sin^2\theta\sin\phi\cos\phi & \sin\theta\cos\theta\cos\phi\\
        \sin^2\theta\sin\phi\cos\phi & \sin^2\theta\sin^2\phi & \sin\theta\cos\theta\sin\phi\\
        \sin\theta\cos\theta\cos\phi & \sin\theta\cos\theta\sin\phi & \cos^2\theta
    \end{pmatrix}\nonumber
\end{equation}
\begin{eqnarray}
   \mathcal{A}_\varepsilon^t&=&\frac{1}{2}\begin{pmatrix}
       2\cos\theta\cos\phi\sin\theta\cos\phi & 2\cos\theta\cos\phi\sin\theta\sin\phi & \cos\theta\cos\phi\cos\theta-\sin^2\theta\cos\phi \\
       2\cos\theta\cos\phi\sin\theta\sin\phi & 2 \cos\theta\sin\phi\sin\theta\sin\phi & \cos\theta\sin\phi\cos\theta-\sin^2\theta\sin\phi\\
      \cos\theta\cos\phi\cos\theta-\sin^2\theta\cos\phi &  \cos\theta\sin\phi\cos\theta-\sin^2\theta\sin\phi & -2\sin\theta\cos\theta
   \end{pmatrix}\nonumber\\
   &=&\begin{pmatrix}
       \frac{1}{2}\sin2\theta\cos^2\phi & \frac{1}{4}\sin2\theta\sin2\phi & \frac{1}{2}\cos2\theta\cos\phi\\
       \frac{1}{4}\sin2\theta\sin2\phi & \frac{1}{2}\sin2\theta\sin^2\phi & \frac{1}{2}\cos2\theta\sin\phi\\
       \frac{1}{2}\cos2\theta\cos\phi & \frac{1}{2}\cos2\theta\sin\phi & -\frac{1}{2}\sin2\theta
   \end{pmatrix}\nonumber
\end{eqnarray}
\begin{eqnarray}
    \mathcal{A}_\varepsilon^w&=&\frac{1}{2}\begin{pmatrix}
        -2\sin\phi\sin\theta\cos\phi & -\sin\phi\sin\theta\sin\phi+\cos\phi\sin\theta\cos\phi & -\sin\phi\cos\theta\\
        -\sin\phi\sin\theta\sin\phi+\cos\phi\sin\theta\cos\phi & 2\cos\phi\sin\theta\sin\phi & \cos\phi\cos\theta\\
        -\sin\phi\cos\theta & \cos\phi\cos\theta & 0
    \end{pmatrix}\nonumber\\
    &=&\frac{1}{2}\begin{pmatrix}
        -\sin2\phi\sin\theta & \cos2\phi\sin\theta & -\sin\phi\cos\theta\\
        \cos2\phi\sin\theta & \sin2\phi\sin\theta & \cos\phi\cos\theta\\
        -\sin\phi\cos\theta & \cos\phi\cos\theta & 0
    \end{pmatrix}
\end{eqnarray}
The total hole-phonon Hamiltonian can be added as per the following equation:
\begin{equation}
    H_{h-ph}^\alpha=\mathcal{D}_{11}^\alpha\mathcal{A}_{\varepsilon,11}^\alpha+\mathcal{D}_{12}^\alpha\mathcal{A}_{\varepsilon,12}^\alpha+\mathcal{D}_{13}^\alpha\mathcal{A}_{\varepsilon,13}^\alpha+\mathcal{D}_{22}^\alpha\mathcal{A}_{\varepsilon,22}^\alpha+\mathcal{D}_{23}^\alpha\mathcal{A}_{\varepsilon,23}^\alpha+\mathcal{D}_{33}^\alpha\mathcal{A}_{\varepsilon,33}^\alpha
\end{equation}
where $\alpha$ is the polarisation index, and $\mathcal{D}$ are the $4\times4$ LK deformation potential matrices. 
\paragraph{l-polarisation:}
Putting in the local strain terms, we can write:

\begin{equation}
        H_{h-ph}^l=i\sqrt{q}\sqrt{\frac{\hbar}{2V_c\rho v_l}}\begin{pmatrix}
            P_\varepsilon^l+Q_\varepsilon^l & 0 & L_\varepsilon^l & M_\varepsilon^l\\
            0 & P_\varepsilon^l+Q_\varepsilon^l & M_\varepsilon^{l^*} & -L_\varepsilon^{l^*}
            \\
            L_\varepsilon^{l^*} & M_\varepsilon^l & P_\varepsilon^l-Q_\varepsilon^l & 0\\
            M_\varepsilon^{l^*} & -L_\varepsilon^l & 0 & P_\varepsilon^l-Q_\varepsilon^l
        \end{pmatrix}e^{i\mathbf{q}\cdot\mathbf{r}}(a_{-\mathbf{q},l}+a^\dagger_{\mathbf{q},l})\nonumber
        \end{equation}
     where,
       \begin{eqnarray}
      P_\varepsilon^l&=&-a(\sin^2\theta\cos^2\phi+\sin^2\theta\sin^2\phi+\cos^2\theta)=-a\nonumber\\
      Q_\varepsilon^l&=&-\frac{b}{2}(\sin^2\theta\cos^2\phi+\sin^2\theta\sin^2\phi- 2\cos^2\theta)=-\frac{b}{2}(1-3\cos^2\theta)\nonumber\\
      L_\varepsilon^l&=&d(\sin\theta\cos\theta\cos\phi-i\sin\theta\cos\theta\sin\phi)=\frac{d}{2}\sin2\theta\,e^{-i\phi}\nonumber\\
      M_\varepsilon^l&=&\frac{\sqrt{3}b}{2}(\sin^2\theta\cos^2\phi-\sin^2\theta\sin^2\phi)-id\sin\phi\cos\phi\sin^2\theta=\frac{\sqrt{3}b}{2}\sin^2\theta\cos2\phi-\frac{id}{2}\sin2\phi\sin^2\theta
      \end{eqnarray}
\paragraph{t-polarisation:}The hole-phonon Hamiltonian in this case takes the following form in angular co-ordinates:
\begin{equation}
        H_{h-ph}^t=i\sqrt{q}\sqrt{\frac{\hbar}{2V_c\rho v_t}}\begin{pmatrix}
            P_\varepsilon^t+Q_\varepsilon^t & 0 & L_\varepsilon^t & M_\varepsilon^t\\
            0 & P_\varepsilon^t+Q_\varepsilon^t & M_\varepsilon^{t^*} & -L_\varepsilon^{t^*}
            \\
            L_\varepsilon^{t^*} & M_\varepsilon^t & P_\varepsilon^t-Q_\varepsilon^t & 0\\
            M_\varepsilon^{t^*} & -L_\varepsilon^t & 0 & P_\varepsilon^t-Q_\varepsilon^t
        \end{pmatrix}e^{i\mathbf{q}\cdot\mathbf{r}}(a_{-\mathbf{q},t}+a^\dagger_{\mathbf{q},t})\nonumber
        \end{equation}
        where,
        \begin{eqnarray}
            P_\varepsilon^t&=&-a\left(\frac{1}{2}\sin2\theta\cos^2\phi+\frac{1}{2}\sin2\theta\sin^2\phi-\frac{1}{2}\sin2\theta\right)=0\nonumber\\
            Q_\varepsilon^t&=&-\frac{b}{2}\left(\frac{1}{2}\sin2\theta\cos^2\phi+\frac{1}{2}\sin2\theta\sin^2\phi+\sin2\theta\right)=-\frac{3b}{4}\sin2\theta\nonumber
        \end{eqnarray}
        \begin{eqnarray}
            L_\varepsilon^t&=&d\left(\frac{1}{2}\cos2\theta\cos\phi-\frac{i}{2}\cos2\theta\sin\phi\right)=\frac{d}{2}\cos2\theta e^{-i\phi}\nonumber\\
            M_\varepsilon^t&=&\frac{\sqrt{3}b}{2}\left(\frac{1}{2}\sin2\theta\cos^2\phi-\frac{1}{2}\sin2\theta\sin^2\phi\right)-\frac{id}{4}\sin2\theta\sin2\phi=\frac{\sqrt{3}b}{4}\sin2\theta\cos2\phi-\frac{id}{4}\sin2\theta\sin2\phi
        \end{eqnarray}
\paragraph{w-polarisation:}For the $3^{\text{rd}}$ polarisation direction:
\begin{equation}
        H_{h-ph}^w=i\sqrt{q}\sqrt{\frac{\hbar}{2V_c\rho v_w}}\begin{pmatrix}
            P_\varepsilon^w+Q_\varepsilon^w & 0 & L_\varepsilon^w & M_\varepsilon^w\\
            0 & P_\varepsilon^w+Q_\varepsilon^w & M_\varepsilon^{w^*} & -L_\varepsilon^{w^*}
            \\
            L_\varepsilon^{w^*} & M_\varepsilon^w & P_\varepsilon^w-Q_\varepsilon^w & 0\\
            M_\varepsilon^{w^*} & -L_\varepsilon^w & 0 & P_\varepsilon^w-Q_\varepsilon^w
        \end{pmatrix}e^{i\mathbf{q}\cdot\mathbf{r}}(a_{-\mathbf{q},w}+a^\dagger_{\mathbf{q},w})\nonumber
        \end{equation}
where,
\begin{eqnarray}
    P_\varepsilon^w&=&-a\left(-\frac{1}{2}\sin2\phi\sin\theta+\frac{1}{2}\sin2\phi\sin\theta\right)=0\nonumber\\
    Q_\varepsilon^w&=&-\frac{b}{2}\left(-\frac{1}{2}\sin2\phi\sin\theta+\frac{1}{2}\sin2\phi\sin\theta\right)=0\nonumber
\end{eqnarray}
\begin{eqnarray}
    L_\varepsilon^w&=&d\left(-\frac{1}{2}\sin\phi\cos\theta-\frac{i}{2}\cos\phi\cos\theta\right)=-\frac{id}{2}\cos\theta e^{-i\phi}\nonumber\\
    M_\varepsilon^w&=& \frac{\sqrt{3}b}{2}\left(-\frac{1}{2}\sin2\phi\sin\theta-\frac{1}{2}\sin2\phi\sin\theta\right)-\frac{id}{2}\cos2\phi\sin\theta=-\frac{\sqrt{3}b}{2}\sin2\phi\sin\theta-\frac{id}{2}\cos2\phi\sin\theta
\end{eqnarray}
The relaxation rate is characterized by the spontaneous and stimulated phonon scattering, hence:
\begin{equation}
    \Gamma_1=\frac{1}{T_1}=\sum_\alpha\left(\frac{2\pi}{\hbar}\sum_{\mathbf{q}}\left|\left\langle\mathbb{0}\right|H_{h-ph}\left|\mathbb{1}\right\rangle_\alpha\right|^2\delta\left(\Delta\mathbb{E}-\hbar\omega_{\alpha,\mathbf{q}}\right)\right)
\end{equation}
The summation over the wave vectors can be changed to continuous integral, and the creation-annihilation operators can be approximated to produce a factor of $N_q$, which denotes the number of acoustic phonons with $q$ momentum:
\begin{eqnarray}\label{seq:reltime}
    \Gamma_1=\frac{1}{T_1}&=&\sum_\alpha\left(\frac{2\pi}{\cancel{\hbar}}\frac{\cancel{V}}{(2\pi)3}\int_{V_q} d^3\mathbf{q}\, q \frac{\cancel{\hbar}}{2\cancel{V_c}\rho v_\alpha}\left|\left\langle\mathbb{0}\right|e^{i\mathbf{q}\cdot\mathbf{r}}H^{LKBP}_{strain,\alpha} \cancel{N_q}\left|\mathbb{1}\right\rangle_\alpha\right|^2\delta\left(\Delta\mathbb{E}-\hbar v_\alpha q\right)\right)\nonumber\\
    &=&\sum_\alpha\left(\frac{1}{8\pi^2}\int_{V_q} q^3\,dq\,\sin\theta\,d\theta\,d\phi \frac{1}{\rho v_\alpha}\left|\left\langle\mathbb{0}\right|e^{i\mathbf{q}\cdot\mathbf{r}}H^{LKBP}_{strain,\alpha} \left|\mathbb{1}\right\rangle_\alpha\right|^2 \frac{1}{\hbar v_\alpha}\delta\left(\frac{\Delta\mathbb{E}}{\hbar v_\alpha}-q\right)\right)\nonumber\\
    &=&\sum_\alpha\left(\frac{1}{8\pi^2\hbar\rho v_\alpha^2}\int_0^{2\pi}\!d\phi\,\int_0^\pi\!d\theta\sin\theta\,\int_0^\infty\!q^3dq\, \left|\left\langle\mathbb{0}\right|e^{i\mathbf{q}\cdot\mathbf{r}}H^{LKBP}_{strain,\alpha} \left|\mathbb{1}\right\rangle_\alpha\right|^2 \delta\left(\frac{\Delta\mathbb{E}}{\hbar v_\alpha}-q\right)\right)
\end{eqnarray}
The qubit ground state $|\mathbb{0}\rangle$ and excited state  $|\mathbb{1}\rangle$ are $1\times4$ spinors with each component multiplied to spatial functions.
\begin{eqnarray}\label{seq:gih-phej}
    &\Rightarrow&\left\langle\mathbb{0}\right|e^{i\mathbf{q}\cdot\mathbf{r}}H^{LKBP}_{strain,\alpha}\left|\mathbb{1}\right\rangle=\begin{pmatrix}
        g_1^* & g_2^* & g_3^* & g_4^*
    \end{pmatrix}e^{i\mathbf{q}\cdot\mathbf{r}}\begin{pmatrix}
    P_\varepsilon^\alpha+Q_\varepsilon^\alpha & 0 & L_\varepsilon^\alpha & M_\varepsilon^\alpha\\
    0 & P_\varepsilon^\alpha+Q_\varepsilon^\alpha & M_\varepsilon^{\alpha^*} & -L_\varepsilon^{\alpha^*} \\
    L_\varepsilon^{\alpha^*} & M_\varepsilon^\alpha & P_\varepsilon^\alpha-Q_\varepsilon^\alpha & 0 \\
    M_\varepsilon^{\alpha^*} & -L_\varepsilon^\alpha & 0 & P_\varepsilon^\alpha-Q_\varepsilon^\alpha \\
    \end{pmatrix}\begin{pmatrix}
        e_1\\
        e_2\\
        e_3\\
        e_4
    \end{pmatrix}\nonumber\\
    &=&\begin{pmatrix}
        g_1^* & g_2^* & g_3^* & g_4^*
    \end{pmatrix}e^{i\mathbf{q}\cdot\mathbf{r}}\begin{pmatrix}
        (P_\varepsilon^\alpha+Q_\varepsilon^\alpha)e_1+L_\varepsilon^\alpha e_3+M_\varepsilon^\alpha e_4\\
        (P_\varepsilon^\alpha+Q_\varepsilon^\alpha)e_2+M_\varepsilon^{\alpha^*} e_3-L_\varepsilon^{\alpha^*} e_4\\
        L_\varepsilon^{\alpha^*} e_1+M_\varepsilon^\alpha e_2+(P_\varepsilon^\alpha-Q_\varepsilon^\alpha)e_3\\
         M_\varepsilon^{\alpha^*} e_1-L_\varepsilon^\alpha e_2+(P_\varepsilon^\alpha-Q_\varepsilon^\alpha)e_4\\
    \end{pmatrix}\nonumber\\
    &=&\left[\left\{(P_\varepsilon^\alpha+Q_\varepsilon^\alpha)e^{i\mathbf{q}\cdot\mathbf{r}}g_1^*e_1+L_\varepsilon^\alpha e^{i\mathbf{q}\cdot\mathbf{r}}g_1^*e_3+M_\varepsilon^\alpha e^{i\mathbf{q}\cdot\mathbf{r}}g_1^*e_4\right\}+\left\{(P_\varepsilon^\alpha+Q_\varepsilon^\alpha)e^{i\mathbf{q}\cdot\mathbf{r}}g_2^*e_2+M_\varepsilon^{\alpha^*} e^{i\mathbf{q}\cdot\mathbf{r}}g_2^*e_3-L_\varepsilon^{\alpha^*} e^{i\mathbf{q}\cdot\mathbf{r}}g_2^*e_4\right\}\right.\nonumber\\
    &\!+\!&\!\left.\left\{L_\varepsilon^{\alpha^*}e^{i\mathbf{q}\cdot\mathbf{r}}g_3^*e_1\!+\!M_\varepsilon^{\alpha} e^{i\mathbf{q}\cdot\mathbf{r}}g_3^*e_2\!+\!(P_\varepsilon^\alpha-Q_\varepsilon^\alpha) e^{i\mathbf{q}\cdot\mathbf{r}}g_3^*e_3\right\}\!+\!\left\{M_\varepsilon^{\alpha^*}e^{i\mathbf{q}\cdot\mathbf{r}}g_4^*e_1\!-\!L_\varepsilon^{\alpha} e^{i\mathbf{q}\cdot\mathbf{r}}g_4^*e_2+\!(P_\varepsilon^\alpha\!-\!Q_\varepsilon^\alpha) e^{i\mathbf{q}\cdot\mathbf{r}}g_4^*e_4\right\}\right]
\end{eqnarray}
According to our model $g_i^*{=}\sum_{\{m,n,l,i'\}}c_{ii'}^{g^*}\psi_n(x)\psi_m(y)\psi_l(z)$, and $e_j^*{=}\sum_{\{m',n',l',j'\}}c^e_{jj'}\psi_{n'}(x)\psi_{m'}(y)\psi_{l'}(z)$; implies that the terms in Eqn.~\ref{seq:gih-phej} have the form
\begin{eqnarray}\label{seq:eqrgeexp}
    e^{i\mathbf{q}\cdot\mathbf{r}}g_i^*e_j\!\Rightarrow\!\left(\sum_{i',j'} c^{g^*}_{ii'}c^e_{jj'}\right)\!\left(\sum_{n,n'}\sum_{m,m'}\sum_{l,l'}\right)\int_{-\infty}^\infty \!dx\,e^{iq_xx}\psi_n(x)\psi_{n'}(x) \int_{-\infty}^\infty \!dy\,e^{iq_yy}\psi_m(y)\psi_{m'}(y) \int_{-\infty}^\infty \!dz\,e^{iq_zz}\psi_l(z)\psi_{l'}(z)\nonumber\\
\end{eqnarray}
To our advantage, the inversion-symmetric basis wave-functions we use to describe our hole QD, i.e. an infinite barrier in $z$ and harmonic potential in $x{-}y$, have closed form solutions of the $e^{i\mathbf{q}\cdot\mathbf{r}}$ integrals. This allows us to evaluate the Relaxation rate $\Gamma_1$ analytically. We also show the dipole approximation to agree with the analytical results for $T_1$; thirdly, a numerical pathway is sketched as an alternative. 
\paragraph*{{\textbf{Analytical: in-plane integrals.}}} The matrix elements of $e^{iq_xx}$ between two $x$(or $y$) wavefunctions are given by,
\begin{equation}
    \left\langle n\right|e^{iq_xx}\left|n'\right\rangle=\int_{-\infty}^\infty \!dx\,e^{iq_xx}\psi_n(x)\psi_{n'}(x)
\end{equation}
$e^{iq_xx}$ can be written as an infinite expansion in Hermite polynomial basis:
\begin{equation}
    e^{iq_xx}=e^{iq_xL_x\frac{x}{L_x}}=e^{\frac{(-iq_xL_x)^2}{4}}\sum_{r=0}^\infty \frac{(iq_xL_x)^r}{2^r r!}H_r\!\left(\frac{x}{L_x}\right)
\end{equation}
\begin{equation}
    \Rightarrow\left\langle n\right|e^{iq_xx}\left|n'\right\rangle=\frac{1}{\sqrt{2^n n! L_x\sqrt{\pi}}}\frac{1}{\sqrt{2^{n'} n'! L_x\sqrt{\pi}}}e^{-\frac{q_x^2L_x^2}{4}}\sum_{r=0}^\infty\frac{(iq_xL_x)^r}{2^r r!}\mathbb{I}(r,n,n')
\end{equation}
with $\mathbb{I}(r,n,n')\!=\!\int_{-\infty}^\infty dx\,e^{-\frac{x^2}{L_x^2}}H_r\!\left(\frac{x}{L_x}\right)H_n\!\left(\frac{x}{L_x}\right)H_{n'}\!\left(\frac{x}{L_x}\right)\!=\!L_x\int_{-\infty}^\infty d\!\left(\frac{x}{L_x}\right)\,e^{-\frac{x^2}{L_x^2}}H_r\!\left(\frac{x}{L_x}\right)H_n\!\left(\frac{x}{L_x}\right)H_{n'}\!\left(\frac{x}{L_x}\right)$. Substituting $\frac{x}{L_x}\rightarrow x$, we write:
\begin{eqnarray}
    \Rightarrow\left\langle n\right|e^{iq_xx}\left|n'\right\rangle&=&\left(2^{n+n'}n!n'!\pi\right)^{-\frac{1}{2}}e^{-\frac{q_x^2L_x^2}{4}}\sum_{r=0}^\infty\frac{(iq_xL_x)^r}{2^r r!}\int_{-\infty}^\infty dx\,e^{-x^2}H_r\!(x)H_n\!(x)H_{n'}\!(x)
\end{eqnarray}
The product of two Hermite polynomials can be expanded in the Hermite polynomial basis:\begin{equation}
    H_n(x)H_{n'}(x)=2^nn!n'!\sum_{k=0}^n\frac{H_{2k+n'-n}(x)}{2^kk!(k+n'-n)!(n-k)!}
\end{equation}
\begin{eqnarray}
    \Rightarrow\!\left\langle n\right|e^{iq_xx}\left|n'\right\rangle\!&=&\!\frac{e^{-\frac{q_x^2L_x^2}{4}}}{\sqrt{2^{n+n'}n!n'!\pi}}\sum_{r=0}^\infty\frac{(iq_xL_x)^r}{2^r r!}2^nn!n'!\sum_{k=0}^n\frac{1}{2^kk!(k+n'-n)!(n-k)!}\int_{-\infty}^\infty dx\,e^{-x^2}H_r\!(x)H_{2k+n'-n}(x)\nonumber\\
    &=&\!\frac{e^{-\frac{q_x^2L_x^2}{4}}}{\sqrt{2^{n+n'}n!n'!\pi}}\sum_{r=0}^\infty\frac{(iq_xL_x)^r}{2^r r!}2^nn!n'!\sum_{k=0}^n\frac{2^{2k+n'-n}(2k+n'-n)!\sqrt{\pi}\delta_{r,2k+n'-n}}{2^kk!(k+n'-n)!(n-k)!}
\end{eqnarray}
where we have used the orthonormality relation:$\int_{-\infty}^\infty e^{-x^2}H_n(x)H_{n'}(x)\,dx=2^nn!\sqrt{\pi}\delta_{n,n'}$. The $\delta$-function boils the $r$-sum down to only one term, such that:
\begin{eqnarray}
    &&\Rightarrow\!\left\langle n\right|e^{iq_xx}\left|n'\right\rangle\!=\!\sqrt{\frac{n'!n!}{2^{n'-n}}}e^{-\frac{q_x^2L_x^2}{4}}\sum_{k=0}^n \frac{(iq_xL_x)^{2k+n'-n}}{\cancel{2^{2k+n'-n}} \cancel{(2k+n'-n)!}}\frac{\cancel{2^{2k+n'-n}}\cancel{(2k+n'-n)!}}{2^kk!(k+n'-n)!(n-k)!}\nonumber\\
    &=&\sqrt{\frac{n!}{2^{n'-n}n'!}}e^{-\frac{q_x^2L_x^2}{4}}\sum_{k=0}^n\frac{n'!(iq_xL_x)^{n'-n}(iq_xL_x)^{2k}}{2^kk!(k+n'-n)!(n-k)!}=\sqrt{\frac{n!}{2^{n'-n}n'!}}e^{-\frac{q_x^2L_x^2}{4}}(iq_xL_x)^{n'-n}\sum_{k=0}^n\frac{n'!((iq_xL_x)^2)^k}{2^kk!(k+n'-n)!(n-k)!}\nonumber\\
    &=&\sqrt{\frac{n!}{2^{n'-n}n'!}}e^{-\frac{q_x^2L_x^2}{4}}(iq_xL_x)^{n'-n}\sum_{k=0}^n\frac{(-1)^k(n+n'-n)!}{k!(k+n'-n)!(n-k)!}\left(\frac{q_x^2L_x^2}{2}\right)^k
\end{eqnarray}
Using the formula for associated Laguerre polynomial $\mathcal{L}_n^a(x)=\sum_{k=0}^n\frac{(-1)^k(n+a)!}{k!(k+a)!(n-k)!}x^k$, the matrix element of $e^{iq_xx}$ can be analytically evaluated as:
\begin{equation}
    \Rightarrow\!\left\langle n\right|e^{iq_xx}\left|n'\right\rangle\!=\!\sqrt{\frac{1}{2^{n'-n}}\frac{n!}{n'!}}e^{-\frac{q_x^2L_x^2}{4}}(iq_xL_x)^{n'-n}\mathcal{L}_n^{n'-n}\!\left(\frac{q_x^2L_x^2}{2}\right)\nonumber
\end{equation}
\begin{equation}\label{seq:eq_yy}
    \Rightarrow\!\left\langle m\right|e^{iq_yy}\left|m'\right\rangle\!=\!\sqrt{\frac{1}{2^{m'-m}}\frac{m!}{m'!}}e^{-\frac{q_y^2L_y^2}{4}}(iq_yL_y)^{m'-m}\mathcal{L}_m^{m'-m}\!\left(\frac{q_y^2L_y^2}{2}\right)
\end{equation}
The $z$-basis wavefunctions are:
\begin{equation}
  \psi_l(z)= \begin{cases}
      \cos\!\left(\frac{(l+1)\pi z}{L_z}\right) & l=0,2,4,..\\
      \sin\!\left(\frac{(l+1)\pi z}{L_z}\right) & l=1,3,5..\\
    \end{cases}\  ,z\in \left\{-\frac{L_z}{2},\frac{L_z}{2}\right\}  
\end{equation}
The matrix element $\left\langle l\right|e^{iq_zz}\left|l'\right\rangle$ is evauated as:
\begin{eqnarray}
&&\int_{-\frac{L_z}{2}}^\frac{L_z}{2}dz\,e^{iq_zz}\psi_l(z)\psi_{l'}(z)=\int_{-\frac{L_z}{2}}^\frac{L_z}{2}dz\,(\cos q_zz+i\sin q_zz)\psi_r(z)\psi_{r'}(z)
\end{eqnarray}
where $r=l+1$ is the new index for z-integral (for simpler mathematical expressions). 
\paragraph*{\underline{$r$ odd, $r'$ even.}} 
For the case of evaluating the matrix element between even and odd z-basis functions is calculated as
\begin{eqnarray}
    &=&\int_{-\frac{L_z}{2}}^\frac{L_z}{2}dz(\cos q_zz+i\sin q_zz)\sqrt{\frac{2}{L_z}}\cos\!\left(\frac{r\pi z}{L_z}\right)\sqrt{\frac{2}{L_z}}\sin\!\left(\frac{r'\pi z}{L_z}\right)\nonumber\\
    &=&\frac{2}{L_z}\int_{-\frac{L_z}{2}}^\frac{L_z}{2}dz(\cos q_zz+i\sin q_zz)\frac{1}{2}\left(\sin\!\left(\frac{\overline{r+r'}\pi z}{L_z}\right)-\sin\!\left(\frac{\overline{r-r'}\pi z}{L_z}\right)\right)\nonumber\\
    &=&\frac{1}{L_z}\int_{-\frac{L_z}{2}}^\frac{L_z}{2}dz\left(\cos q_zz\sin\!\left(\frac{\overline{r+r'}\pi z}{L_z}\!\right)\!-\cos q_zz\sin\!\left(\frac{\overline{r-r'}\pi z}{L_z}\!\right)\!+i\sin q_zz\sin\!\left(\frac{\overline{r+r'}\pi z}{L_z}\!\right)\!-i\sin q_zz\sin\!\left(\frac{\overline{r-r'}\pi z}{L_z}\right)\right)\nonumber\\
    &=&\frac{1}{2L_z}\int_{-\frac{L_z}{2}}^\frac{L_z}{2}dz\left(\sin\left(\! q_zz+\frac{\overline{r+r'}\pi z}{L_z}\right)\!-\sin\left(\!q_zz-\frac{\overline{r+r'}\pi z}{L_z}\right)\!-\sin\left(\!q_zz+\frac{\overline{r-r'}\pi z}{L_z}\right)\!+\sin\left(\!q_zz-\frac{\overline{r-r'}\pi z}{L_z}\right)\!\right)\nonumber\\
    &+&\frac{i}{2L_z}\int_{-\frac{L_z}{2}}^\frac{L_z}{2}dz\left(\!\cos\!\left(\! q_zz+\frac{\overline{r+r'}\pi z}{L_z}\right)\!+\cos\!\left(\!q_zz-\frac{\overline{r+r'}\pi z}{L_z}\right)\!-\cos\!\left(\!q_zz+\frac{\overline{r-r'}\pi z}{L_z}\right)\!-\sin\!\left(\!q_zz-\frac{\overline{r-r'}\pi z}{L_z}\right)\!\right)
\end{eqnarray}
The $\sin$ and $\cos$ terms can be now evaluated straight forwardly to give us the final simplified expression:
\begin{equation}
    =\frac{i}{2}\left[\frac{\sin\!\left(\frac{q_zL_z}{2}+\frac{(r+r')\pi}{2}\right)}{\frac{q_zL_z}{2}+\frac{(r+r')\pi}{2}}-\frac{\sin\!\left(\frac{q_zL_z}{2}-\frac{(r+r')\pi}{2}\right)}{\frac{q_zL_z}{2}-\frac{(r+r')\pi}{2}}-\frac{\sin\!\left(\frac{q_zL_z}{2}+\frac{(r-r')\pi}{2}\right)}{\frac{q_zL_z}{2}+\frac{(r-r')\pi}{2}}+\frac{\sin\!\left(\frac{q_zL_z}{2}-\frac{(r-r')\pi}{2}\right)}{\frac{q_zL_z}{2}-\frac{(r-r')\pi}{2}}\right]
\end{equation}
\paragraph*{\underline{$r,r'$ both odd/even.}} 
For the case of evaluating the matrix element between both even(both odd) z-basis functions is calculated as
\begin{eqnarray}\label{seq:eq_zz}
    &=&\int_{-\frac{L_z}{2}}^\frac{L_z}{2}dz(\cos q_zz+i\sin q_zz)\sqrt{\frac{2}{L_z}}\sin\!\left(\frac{r\pi z}{L_z}\right)\sqrt{\frac{2}{L_z}}\sin\!\left(\frac{r'\pi z}{L_z}\right)\nonumber\\
    &=&-\frac{1}{2}\left[\frac{\sin\!\left(\frac{q_zL_z}{2}+\frac{(r+r')\pi}{2}\right)}{\frac{q_zL_z}{2}+\frac{(r+r')\pi}{2}}+\frac{\sin\!\left(\frac{q_zL_z}{2}-\frac{(r+r')\pi}{2}\right)}{\frac{q_zL_z}{2}-\frac{(r+r')\pi}{2}}+\frac{\sin\!\left(\frac{q_zL_z}{2}+\frac{(r-r')\pi}{2}\right)}{\frac{q_zL_z}{2}+\frac{(r-r')\pi}{2}}+\frac{\sin\!\left(\frac{q_zL_z}{2}-\frac{(r-r')\pi}{2}\right)}{\frac{q_zL_z}{2}-\frac{(r-r')\pi}{2}}\right]
\end{eqnarray}
Finally we put the results from Eqns.~\ref{seq:eq_yy}, \ref{seq:eq_zz} into Eqn.~\ref{seq:eqrgeexp} to evaluate:
\begin{eqnarray}
    e^{i\mathbf{q}\cdot\mathbf{r}}g_i^*e_j\!&\Rightarrow&\!\left(\sum_{i',j'} c^{g^*}_{ii'}c^e_{jj'}\right)\!\left(\sum_{n,n'}\sum_{m,m'}\sum_{l,l'}\right)\sqrt{\frac{1}{2^{n'-n}}\frac{n!}{n'!}}e^{-\frac{q_x^2L_x^2}{4}}(iq_xL_x)^{n'-n}\mathcal{L}_n^{n'-n}\!\left(\frac{q_x^2L_x^2}{2}\right)\nonumber\\
    &\times&\sqrt{\frac{1}{2^{m'-m}}\frac{m!}{m'!}}e^{-\frac{q_y^2L_y^2}{4}}(iq_yL_y)^{m'-m}\mathcal{L}_m^{m'-m}\!\left(\frac{q_y^2L_y^2}{2}\right) \times f(l,l',q_zL_z)\nonumber\\
\end{eqnarray}
where,
\begin{equation}
    f(l,l',q_zL_z)=\begin{cases}
    \frac{i}{2}\left[\frac{\sin\!\left(\frac{q_zL_z}{2}+\frac{(r+r')\pi}{2}\right)}{\frac{q_zL_z}{2}+\frac{(r+r')\pi}{2}}-\frac{\sin\!\left(\frac{q_zL_z}{2}-\frac{(r+r')\pi}{2}\right)}{\frac{q_zL_z}{2}-\frac{(r+r')\pi}{2}}-\frac{\sin\!\left(\frac{q_zL_z}{2}+\frac{(r-r')\pi}{2}\right)}{\frac{q_zL_z}{2}+\frac{(r-r')\pi}{2}}+\frac{\sin\!\left(\frac{q_zL_z}{2}-\frac{(r-r')\pi}{2}\right)}{\frac{q_zL_z}{2}-\frac{(r-r')\pi}{2}}\right] & l\pm l'=\text{odd}\\
        -\frac{1}{2}\left[\frac{\sin\!\left(\frac{q_zL_z}{2}+\frac{(r+r')\pi}{2}\right)}{\frac{q_zL_z}{2}+\frac{(r+r')\pi}{2}}+\frac{\sin\!\left(\frac{q_zL_z}{2}-\frac{(r+r')\pi}{2}\right)}{\frac{q_zL_z}{2}-\frac{(r+r')\pi}{2}}+\frac{\sin\!\left(\frac{q_zL_z}{2}+\frac{(r-r')\pi}{2}\right)}{\frac{q_zL_z}{2}+\frac{(r-r')\pi}{2}}+\frac{\sin\!\left(\frac{q_zL_z}{2}-\frac{(r-r')\pi}{2}\right)}{\frac{q_zL_z}{2}-\frac{(r-r')\pi}{2}}\right] & l\pm l'=\text{even}\\
    \end{cases}
\end{equation}
Next we make the substitution $q_x\rightarrow q\sin\theta\cos\phi$, $q_y\rightarrow q\sin\theta\sin\phi$, $q_z\rightarrow q\cos\theta$ to write
\begin{eqnarray}
e^{i\mathbf{q}\cdot\mathbf{r}}g_i^*e_j\!&\Rightarrow&\!\left(\sum_{i',j'} c^{g^*}_{ii'}c^e_{jj'}\right)\!\left(\sum_{n,n'}\sum_{m,m'}\sum_{l,l'}\right)\sqrt{\frac{1}{2^{n'-n}}\frac{n!}{n'!}}e^{-\frac{q^2L_x^2\sin^2\theta\cos^2\phi}{4}}(iqL_x\sin\theta\cos\phi)^{n'-n}\mathcal{L}_n^{n'-n}\!\left(\frac{q^2L_x^2\sin^2\theta\cos^2\phi}{2}\right)\nonumber\\
    &\times&\sqrt{\frac{1}{2^{m'-m}}\frac{m!}{m'!}}e^{-\frac{q^2L_y^2\sin^2\theta\sin^2\phi}{4}}(iqL_y\sin\theta\sin\phi)^{m'-m}\mathcal{L}_m^{m'-m}\!\left(\frac{q^2L_y^2\sin^2\theta\sin^2\phi}{2}\right) \times f(l,l',qL_z\cos\theta)\nonumber\\
\end{eqnarray}
These set of equations allow us to calculate the terms in Eqn.~\ref{seq:gih-phej} to express $\left|\left\langle\mathbb{0}\right|e^{i\mathbf{q}\cdot\mathbf{r}}H^{LKBP}_{strain,\alpha}\left|\mathbb{1}\right\rangle\right|^2$ as $I_\alpha(q,\theta,\phi)$. From Eqn.~\ref{seq:reltime},
\begin{eqnarray}
    \Gamma_1=\frac{1}{T_1}&=&\sum_\alpha\left(\frac{1}{8\pi^2\hbar\rho v_\alpha^2}\int_0^{2\pi}\!d\phi\,\int_0^\pi\!d\theta\sin\theta\,\int_0^\infty\!q^3dq\, I_\alpha(q,\theta,\phi)\, \delta\!\left(\frac{\Delta\mathbb{E}}{\hbar v_\alpha}-q\right)\right)\nonumber\\
    &=&\sum_\alpha\left(\frac{\Delta\mathbb{E}^3}{8\pi^2\hbar^4\rho v_\alpha^5}\int_0^{2\pi}\!d\phi\,\int_0^\pi\!d\theta\sin\theta\,I_\alpha\!\left(q\rightarrow \frac{\Delta\mathbb{E}}{\hbar v_\alpha},\theta,\phi\right)\right)=\sum_\alpha\left(\frac{\Delta\mathbb{E}^3}{8\pi^2\hbar^4\rho v_\alpha^5}\,\Omega_\alpha\!\left(\frac{\Delta\mathbb{E}}{\hbar v_\alpha}\right)\right)
\end{eqnarray}
where $\Omega$ denotes the angular integration.
\paragraph*{{\textbf{Dipole approximation.}}} Alternatively, one can expand $e^{i\mathbf{q}\cdot\mathbf{r}}\approx1(+i\mathbf{q}\cdot\mathbf{r}-|\mathbf{q}\cdot\mathbf{r}|^2+..)$ which simplifies the product $\left|\left\langle\mathbb{0}\right|e^{i\mathbf{q}\cdot\mathbf{r}}H^{LKBP}_{strain,\alpha}\left|\mathbb{1}\right\rangle\right|^2\approx\left|\left\langle\mathbb{0}\right|H^{LKBP}_{strain,\alpha}\left|\mathbb{1}\right\rangle+i\left\langle\mathbb{0}\right|\mathbf{q}\cdot\mathbf{r}H^{LKBP}_{strain,\alpha}\left|\mathbb{1}\right\rangle-\left\langle\mathbb{0}\right||\mathbf{q}\cdot\mathbf{r}|^2H^{LKBP}_{strain,\alpha}\left|\mathbb{1}\right\rangle+..\right|^2$. In sharp contrast to electron spin-$\frac{1}{2}$ qubits, where the local strain is a diagonal tensor and the leading zeroth order term $\left\langle\mathbb{0}\left|H_\varepsilon\right|\mathbb{1}\right\rangle$ vanishes; for the spin-$\frac{3}{2}$ holes the leading term is the zeroth order, resulting in a $\alpha_r B^3+\beta_r B^4 + \gamma_r B^5$ relaxation rate variation, compared to $B^5$ variation in electron spin-$\frac{1}{2}$ qubits. While the $B^3$ and $B^5$ dependence are explained by the first two terms in dipole approximation, the orbital B-terms in the qubit admixture give rise to the $B^4$ dependence.

\section{Random Telegraph Noise(RTN) Dephasing: Screened Potential of a charge defect}\label{append:dephasing}
Taking into account the screening effect of the 2DHG being formed in Ge, the Fourier $q$-space form of the potential of single defect with charge $e$ is given by:
\begin{equation}
U_{scr}(q)=\frac{e^2}{2 \epsilon_0 \epsilon_r} e^{-q d}\frac{\Theta\left(2 k_F-q\right)}{q+q_{T F}}.
\end{equation}
where $U_{scr}(q)$ is known as the Thomas-Fermi screened potential, $q$ is the Fourier space variable, $q_{\text{TF}}{=}0.49\, \text{nm}^{-1}$ is the Thomas-Fermi wave vector in germanium independent of the density of holes. $k_F$ is the Fermi wave vector which is estimated to be $0.1 \text{nm}^{-1}$ in our calculations, and $\Theta$ is the Heaviside Theta step function. Considering constant screening, which only holds for $q{<}2k_F$, the screened potential in real space under the limit $ d {\ll}r$ is approximated to be:
\begin{equation}
U_{s}(\bm{r})=\frac{e^2}{4 \pi \epsilon_0 \epsilon_r} \frac{1}{q_{\text{TF}}^2}\left( \frac{1}{\left|\mathbf{r}-\mathbf{r}_D\right|^3}+ d\frac{q_{\text{TF}}}{\left|\mathbf{r}-\mathbf{r}_D\right|^3 }\right)
\end{equation}
The relative electrical permeability of germanium is $\epsilon_r = 15.8$; $\epsilon_{0}$ being the vacuum electrical permeability. $\mathbf{r}_D{=}\{80\,\text{nm},80\,\text{nm},0\}$ is the vector denoting the distance of the in-plane charge defect from the center of the QD.

\section{\texorpdfstring{$\bm{g}$}{g}-factor anisotropy and fitting parameters}\label{append:anisotropy}
In the main text, Fig.\ref{fig:Experiment}, we mentioned that the fitting paramters are not unique. Here we suggest some other possible configurations to fit the parameters. The main fitting parameters are dot size ($L_x, L_y, L_z$), the electric field can be tuned via gate to modulate the $g$-factors.
\begin{enumerate}
    \item $L_x$=40 \, nm, $L_y$=60\, nm $L_z$ = 10\,nm
    \item $L_x$=30 \, nm, $L_y$=50\, nm $L_z$ = 10.5\,nm
    \item $L_x$=44 \, nm, $L_y$=60\, nm $L_z$ = 9.5\,nm
\end{enumerate}
Note that there are many other possible combinations of parameters if the strains (both uniaxial strain and shear strain) are included.
\end{widetext}

\end{document}